\documentclass[letterpaper]{aa700}  
\usepackage{graphicx}
\usepackage{txfonts}
\usepackage{lscape}
\usepackage{longtable}
\usepackage{natbib}
\bibpunct{(}{)}{;}{a}{}{,}
\begin{document}
   \title{Fundamentals of the dwarf fundamental plane
          \thanks{Based on observations acquired from CFHT, CTIO, ESO, OAN-SPM, and SAAO}
         }

      \author{Marshall L. McCall\inst{1}
          \and
              O. Vaduvescu\inst{2,3}
          \and
              F. Pozo Nunez\inst{3}
          \and
              A. Barr Dominguez\inst{3}
          \and
              R. Fingerhut\inst{1}
          \and
              E. Unda-Sanzana\inst{3}
          \and 
              Bintao Li\inst{1}
          \and
              M. Albrecht\inst{4,3}
              }

   \offprints{M. L. McCall}

\institute{
York University, Department of Physics and Astronomy, 
4700 Keele Street, Toronto, ON, Canada M3J1P3\\
\email{mccall@yorku.ca}\\
\and
Isaac Newton Group of Telescopes, 
Ap. de correos 321, E-38700 Santa Cruz de la Palma, Canary Islands, Spain\\
\email{ovidiuv@ing.iac.es}\\
\and
Instituto de Astronom\'ia, Universidad Cat\'olica del Norte,
Avenida Angamos 0610, Antofagasta, Chile\\
\and
Argelander-Institut f\"{u}r Astronomie, 
Universit\"{a}t Bonn, Auf dem H\"{u}gel 71, 53121 Bonn, Germany\\
}

   \date{Received 08 July 2011; Accepted 27 January 2012}

 
\abstract
{}
{Star-forming dwarfs are studied to elucidate the physical underpinnings of their fundamental plane.  Processes controlling dynamics are evaluated, connections between quiescent and bursting dwarfs are examined, and the viability of using structural properties of dwarfs to determine distances is assessed.}
{Deep surface photometry in $K_s$ is presented for 19 star-forming dwarfs.  The data are amalgamated with previously published observations to create a sample of 66 galaxies suitable for exploring how global properties and kinematics are connected.}  
{It is confirmed that residuals in the Tully-Fisher relation are correlated with surface brightness, but that even after accommodating the surface brightness dependence through the dwarf fundamental plane, residuals in absolute magnitude are far larger than expected from observational errors.  Rather, a \hbox{\it more} fundamental plane is identified which connects the potential to HI line width and surface brightness.  Residuals correlate with the axis ratio in a way which can be accommodated by recognizing the galaxies to be oblate spheroids viewed at varying angles.  Correction of surface brightnesses to face-on leads to a correlation among the potential, line width, and surface brightness for which residuals are entirely attributable to observational uncertainties.  The mean mass-to-light ratio of the diffuse component of the galaxies is constrained to be $0.88 \pm 0.20$ in $K_s$.  Blue compact dwarfs lie in the same plane as dwarf irregulars.  The dependence of the potential on line width is less strong than expected for virialized systems, but this may be because surface brightness is acting as a proxy for variations in the mass-to-light ratio from galaxy to galaxy.  Altogether, the observations suggest that gas motions are predominantly disordered and isotropic, that they are a consequence of gravity, not turbulence, and that the mass and scale of dark matter haloes scale with the amount and distribution of luminous matter.  The tight relationship between the potential and observables offers the promise of determining distances to unresolved star-forming dwarfs to an accuracy comparable to that provided by the Tully-Fisher relation for spirals.}
{}


   \authorrunning{McCall et al}
   \maketitle
%

\section{Introduction}

$\Lambda$CDM cosmology leads to dwarf galaxies with dark matter haloes at the centre of which is a cusp in density.  However, dwarfs in which a significant portion of their internal energy is ordered show rotation curves which rise less steeply outward from their centres than expected.  Thus, their core density profiles must be quite flat \citep[][and references therein]{moo94a,flo94a,sim05a,blo08a,oh11a}.  Dwarfs over a wide range of absolute magnitudes display surface brightness profiles which are flat in the cores, also suggestive of a central dark matter framework whose density is slowly varying \citep{vad05a}.  Hydrodynamic simulations taking into account star formation and its consequences suggest that a flat density profile is a response to the blowout of baryonic matter with low angular momentum by supernova explosions \citep{gov10a}.  On the contrary, the chemical properties of star-forming dwarfs in low-density environments indicate that gas flows have not played a major role in their evolution \citep{lee03a,vad07a}.
 
A clue as to whether or not the current state of dwarfs is a consequence of gas flows may come from velocity dispersions.  Galaxies whose evolution has been affected significantly by flows may well display internal motions which are not entirely explainable as a response to gravity.  Recently, rotating disk galaxies were discovered at low redshift in which the velocity dispersion is large \citep{gre10a}.  Because line widths correlate with star formation rates but not masses or gas fractions, the unusual motions were attributed to turbulence resulting from star formation activity.  This may be relevant to understanding the Tully-Fisher relation for star-forming dwarfs, which is highly scattered \citep{vad08a}.  Although some of the scatter can be explained through a connection to surface brightness (the fundamental plane for dwarfs), there remains a significant component which cannot be attributed to observational errors \citep{vad08a}.  

Within the context of evaluating the impact of star formation on dynamics, and by implication the evolution of both mass and chemistry, it is important to examine in more detail how closely the mass and distribution of visible matter in dwarfs are linked to kinematics.   This motivates, in particular, exploration of the baryonic Tully-Fisher relation \citep{mcg00a}, since a significant portion of the mass of star-forming dwarfs is in gaseous form.  From the standpoint of turbulence, it is of interest to compare blue compact dwarfs, in which there is evidence for a recent burst of star formation, with the more quiescent dwarf irregular galaxies \citep[e.g.,][]{vad06a}.  A better understanding of the physics of star-forming dwarfs also has the potential to open up new avenues for determining distances.  At the moment, distances to unresolved systems are so poorly constrained that it is not possible to map peculiar motions on large scales independently from giants.

In this paper, star-forming dwarfs in the Local Volume whose structural properties are defined by near-infrared surface photometry are employed to study how the luminosity, baryonic mass, and baryonic potential are linked to kinematics.  Simultaneously, the mass-to-light ratio is constrained by optimizing linkages.  Section~\ref{sec_observations} introduces new near-infrared observations of star-forming dwarfs, the surface photometry for which is presented in Section~\ref{sec_surface_photometry}.  An expanded sample of galaxies suitable for study is assembled in Section~\ref{sec_sample}, and then subjected to detailed analysis in Section~\ref{sec_analysis}.  This leads to the identification of a {\it more} fundamental plane for dwarfs.  Section~\ref{sec_discussion} follows with a discussion of results, especially examining how closely internal motions are tied to gravity.  As well, a new method for deriving distances to dwarfs is presented.  Finally, conclusions are presented in Section~\ref{sec_conclusions}.


\section{Observations}
\label{sec_observations}

\subsection{Blanco observations 2008}

During 2008 Mar 10--13 and Aug 10--12, deep NIR 
images of 23 galaxies were acquired using the $4.1 \, \rm m$ Blanco telescope at Cerro Tololo Inter-American Observatory, Chile 
(Run IDs: 2008A-0913 and 2008B-0909).  All three nights 
of the first run were photometric, but only the second night of the August run was clear.  During both runs, the ISPI camera was used at the $f/8$ Cassegrain focus.  The detector was a Hawaii array with $2048 \times 2048$ pixels.  The scale was $0\,\farcs3 \, \rm pix^{-1}$, yielding a field of view $10\farcm25 
\times 10\farcm25$. Targets were imaged exclusively through the $K_s$ filter.  Table~\ref{tbl_obs_log} summarizes the observations.

To sample the sky, small objects were cycled through four quadrants of the array.  For large targets, the telescope was jogged to a sky field after every pair of dithered target images.  Data were reduced, calibrated, and analyzed in the manner described by \citet{vad05a}.  Typically, 10 to 15 2MASS stars were employed to calibrate each field. 
Imaging and surface photometry for the 13 dwarfs clearly detected with the Blanco telescope are presented in Figure~\ref{fig_photometry}. 

For reference, Figure~\ref{fig_nophotometry} gives the reduced images of the fields of the 10 unexaminable dwarfs.  It is possible that galaxies imaged in August (HIPASS~J1337$-$39, Sag~DIG, and DDO~210) were obscured by thin clouds.  HIPASS~J1351$-$47 and Sag~DIG appear to have been detected, but not well enough to permit surface photometry.  The remaining galaxies were just too faint to detect with the chosen exposures.

\begin{figure*}[tbp]
\centering
\includegraphics[angle=0,width=11.6cm]{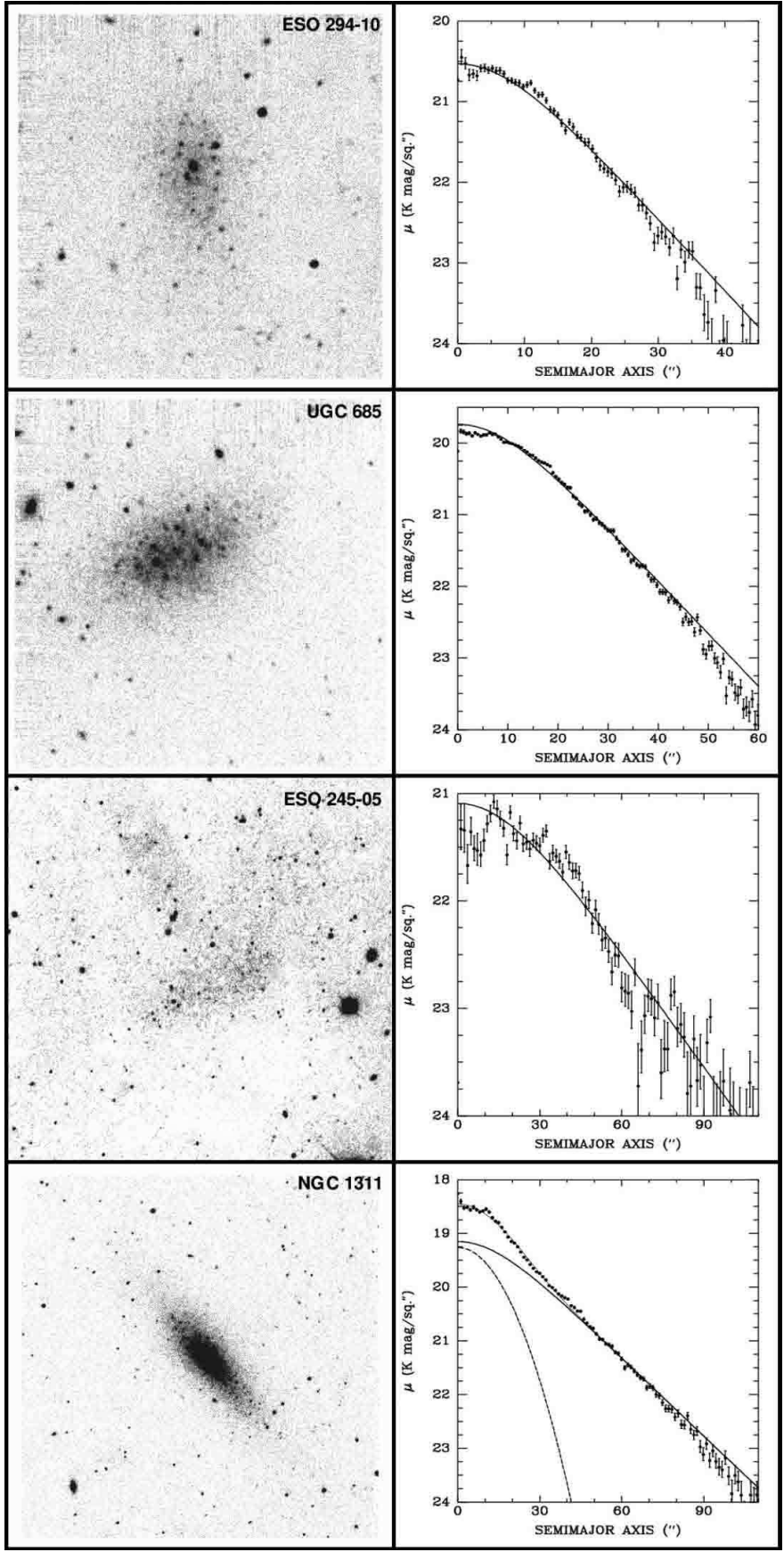}
\caption{Images and surface photometry of dIs observed at CTIO (Blanco) and La Silla (NTT). 
Left panels: $K_s$ images (North is up, East to the left). The field of view is about $5\arcmin \times 5\arcmin$ 
(Blanco) or $2\farcm5 \times 2\farcm5$ (NTT). 
Right panels: Surface brightness profiles in $K_s$ for the unresolved components. 
The thick solid curves are fits of a sech function. In a few cases, a Gaussian burst was fitted simultaneously, 
and is marked by a dashed curve.  In these cases, the sum of the sech and Gaussian components is shown as a thin solid line (sometimes hard to see due to overlap with the observations). 
} 
\label{fig_photometry}
\end{figure*}

\addtocounter{figure}{-1}
\begin{figure*}[tbp]
\centering
\includegraphics[angle=0,width=11.6cm]{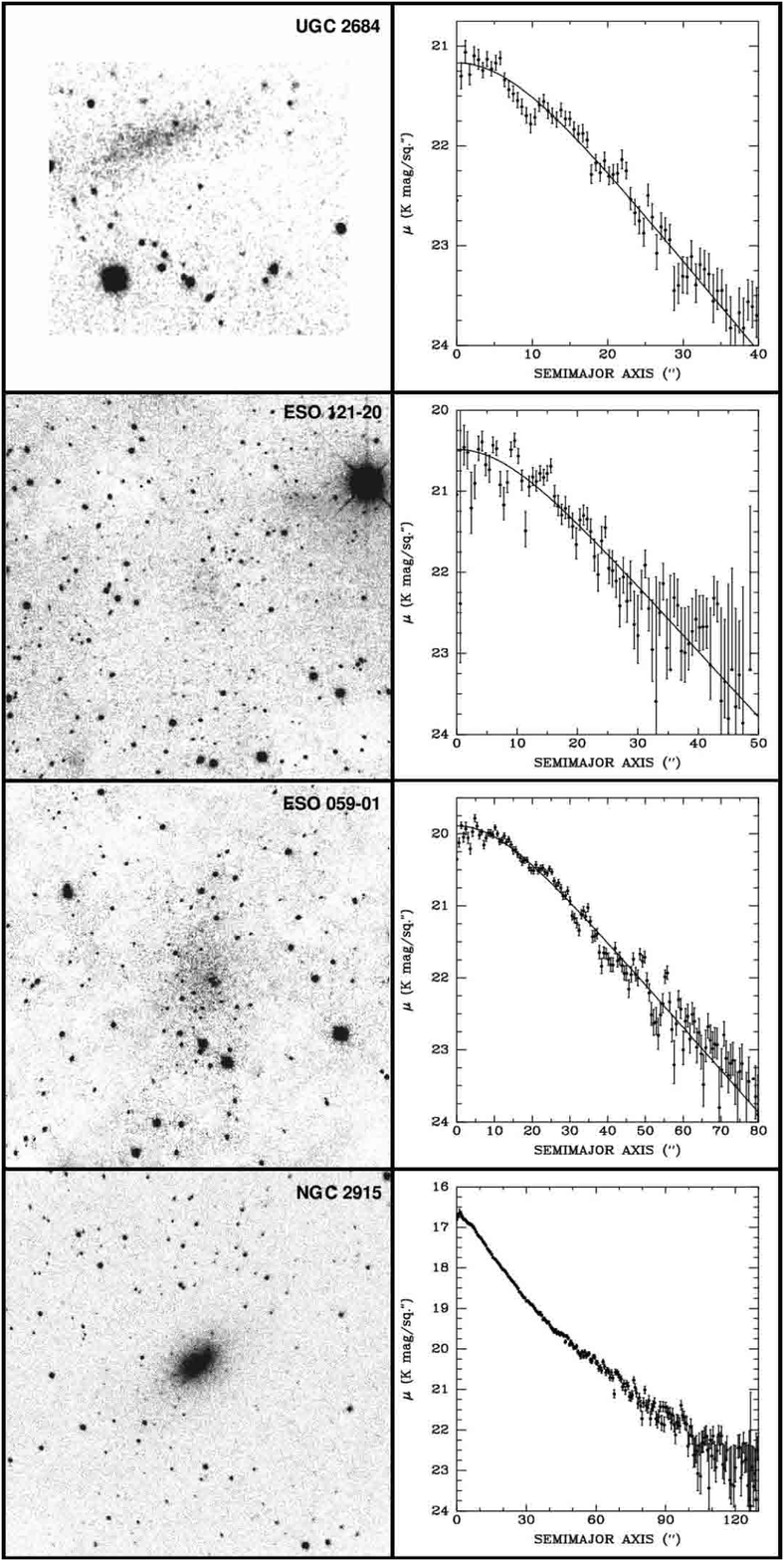}
\caption{
(cont'd)
}
\end{figure*}

\addtocounter{figure}{-1}
\begin{figure*}[tbp]
\centering
\includegraphics[angle=0,width=11.6cm]{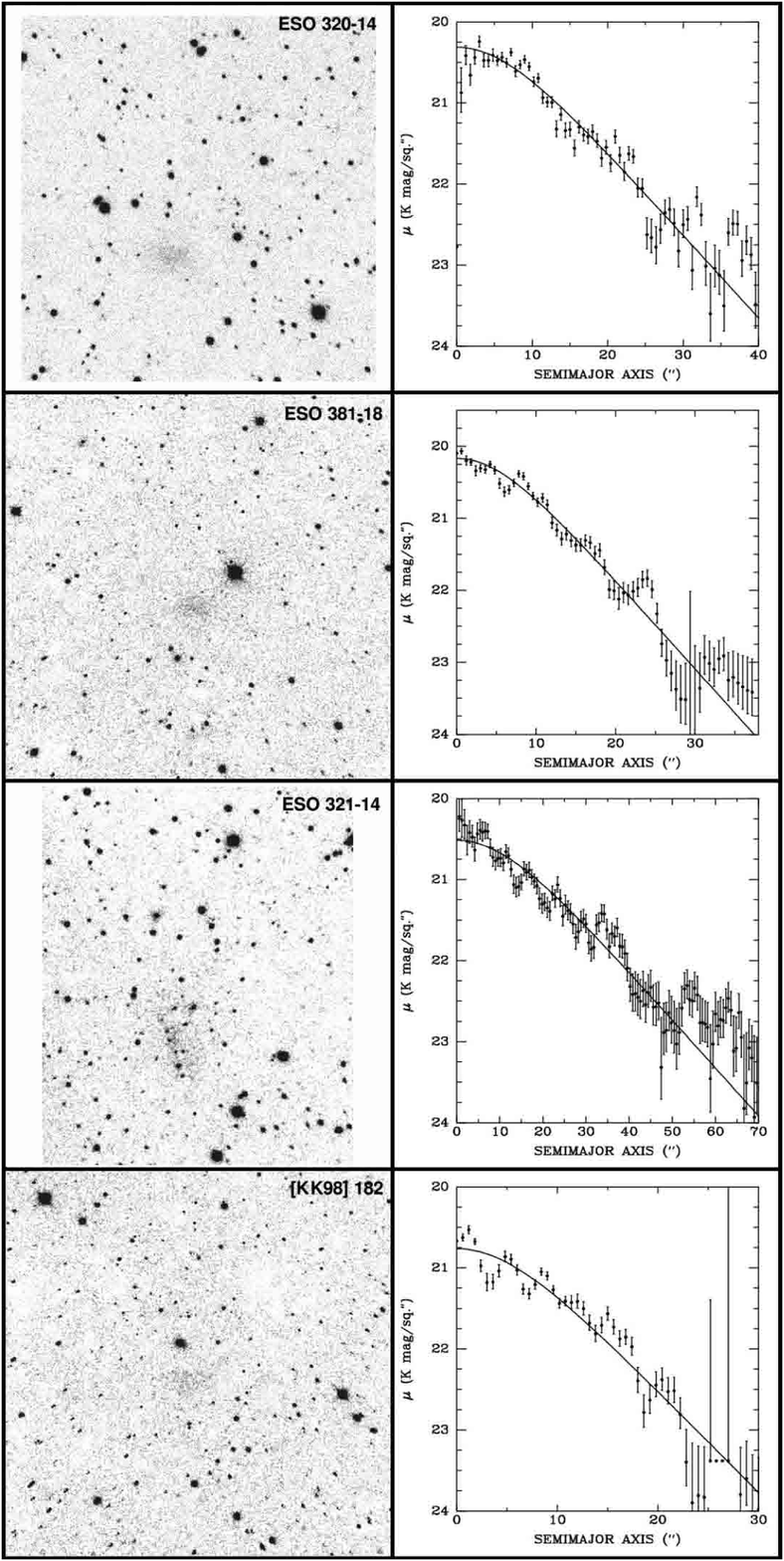}
\caption{
(cont'd)
}
\end{figure*}

\addtocounter{figure}{-1}
\begin{figure*}[tbp]
\centering
\includegraphics[angle=0,width=11.6cm]{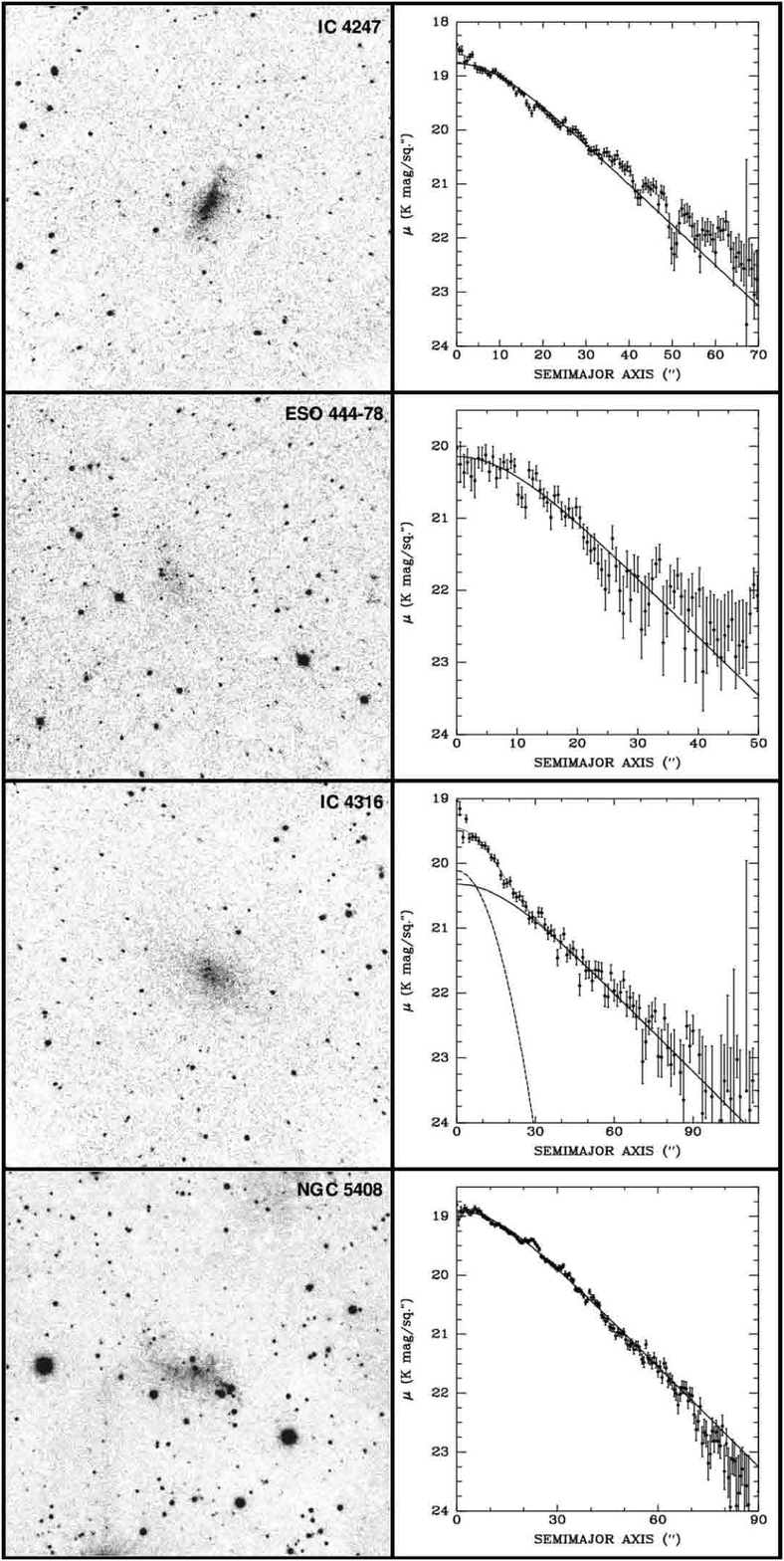}
\caption{
(cont'd)
}
\end{figure*}

\addtocounter{figure}{-1}
\begin{figure*}[tbp]
\centering
\includegraphics[angle=0,width=11.6cm]{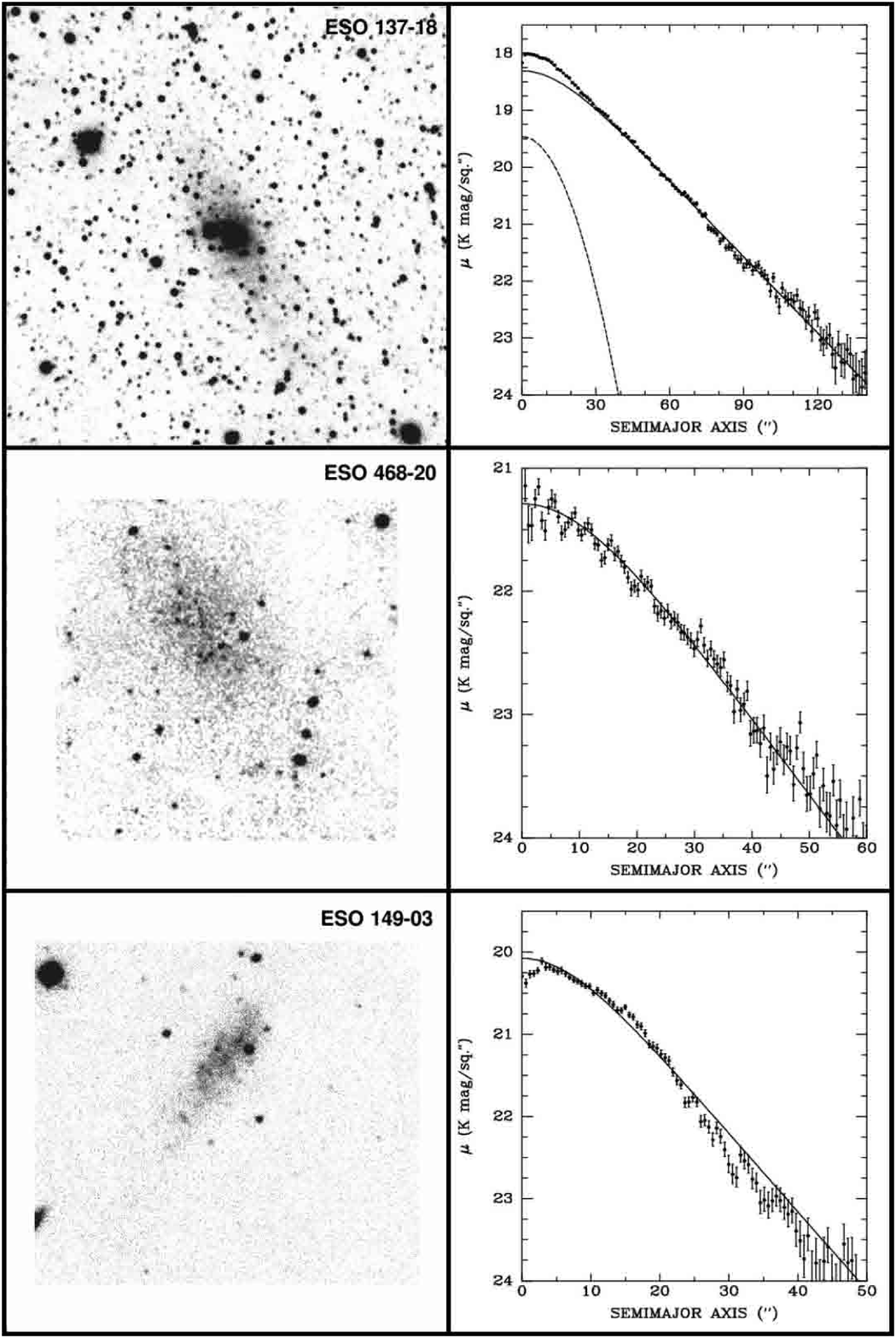}
\caption{
(cont'd)
}
\end{figure*}

\begin{figure*}[tbp]
\centering
\includegraphics[angle=0,width=17cm]{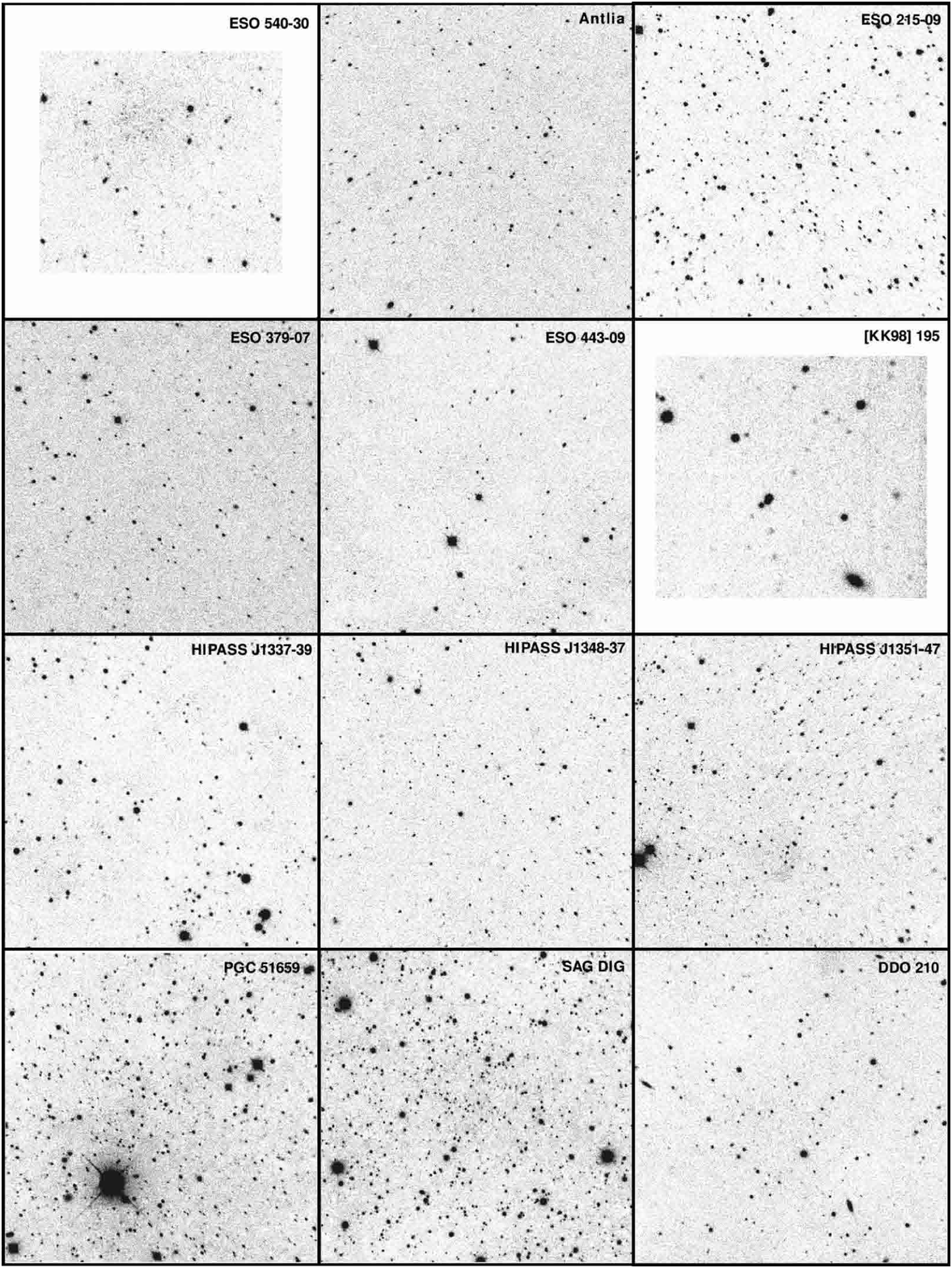}
\caption{Images in $K_s$ of the fields of galaxies either marginally or not detected at CTIO (Blanco) and La Silla (NTT). North is up, and East is to the left. 
The fields of view are about $5\arcmin \times 5\arcmin$ (Blanco) and $2\farcm5 \times 2\farcm5$ (NTT). } 
\label{fig_nophotometry}
\end{figure*} 

\subsection{NTT observations 2008}

During 2008 Aug 13--17, deep NIR imaging of nine galaxies 
was undertaken with the 
3.5m NTT telescope at ESO La Silla Observatory, Chile (Run ID: 081.B-0386(A)).  One night was clear, and the rest were clouded out.  The SOFI camera equipped with a Hawaii HgCdTe array was employed at the $f/11$ Nasmyth focus.  The array was composed of $1024 \times 1024$ pixels.  The scale was $0\,\farcs288 \, \rm pix^{-1}$, so the field of view was $4\farcm92 
\times 4\farcm92$.   All targets were observed with the $K_s$ filter only.   Observations are summarized in Table~\ref{tbl_obs_log}.

Data were reduced, calibrated, and analyzed in the same manner as for the Blanco runs.  Around five 2MASS stars were employed 
to calibrate most of the fields. 
Imaging and surface photometry for the six dIs solidly detected are included in Figure~\ref{fig_photometry}. Figure~\ref{fig_nophotometry} includes the reduced images of the fields of the two unexaminable dwarfs not observed with the Blanco telescope.  In fact,  ESO~540-30 appears to have been detected, but not well enough to carry out surface photometry.

\subsection{Other observations 2002--2007}

As part of separate studies,  deep $K_s$ images of 110 different dwarf galaxies were obtained from 2002 to 2007 over 
observing runs conducted with the 2.1 m telescope of OAN-SPM in Mexico (2002 and 2005), the 1.4m IRSF telescope of SAAO in South Africa (2005 and 2006), the 3.6m CFHT 
in Hawaii (2002, 2004, 2005, and 2006), and the Blanco Telescope at CTIO (2006 and 2007). Images and surface photometry are presented in \cite{vad05a} (34 galaxies), \cite{vad08a} (17 galaxies, plus eight from 2MASS), and \cite{fin10a} (80 galaxies).

\section{Surface photometry}
\label{sec_surface_photometry}

Of the newly-observed dwarfs, 15 out of 19 have flat cores and exponential wings.  \citet{vad05a} showed that a sech function provides a good fit to such profiles.  
For this function, the apparent surface brightness $\mu^{app}$ in $\rm mag \, arcsec^{-2}$ at radius $r$ along the major axis is given by
\begin{equation}
\mu^{app} = \mu_0^{app} - 2.5 \log {2 \over e^{r / r_0} +e^{-r / r_0}} 
\end{equation}
where $\mu_{0}^{app}$ is the apparent surface brightness at the centre and $r_0$ is the scale length.  Solutions for the parameters of the best fitting sech functions are given in Table~\ref{tbl_measurements}.
In the right panels of Figure~\ref{fig_photometry}, fits are shown as thick solid lines.

Some dwarfs show an excess of light in their centres.  They can be interpreted as being normal dIs hosting a central starburst, i.e., blue compact dwarfs \citep{vad06a}.   NGC~1311 and ESO~137-18 are two such objects.  Their surface brightness profiles were modeled  by simultaneously fitting a Gaussian on top of the sech function describing the extended underlying light distribution.  In Figure~\ref{fig_photometry}, the Gaussian component is displayed as a dashed line, and the sum of the Gaussian and sech functions is marked by a thin solid line.

The brightness of the main body of each dwarf was estimated by integrating the sech function out to infinity.  The apparent total magnitude $m_{sech}^{app}$, referred to here as the {\it sech magnitude}, was computed from
\begin{equation}
m_{sech}^{app} = -2.5 \log \left[ 11.51036 \, r_0^2 \, q \, I_0^{app}  \right]  
\end{equation}
where $I_0^{app}$ is the apparent central surface brightness in linear units and $q$ is the axis ratio ($b/a$) of the isophotes.   The magnitude within the outermost detected isophote, referred to here as the {\it isophotal magnitude}, was also estimated.  Sech and isophotal magnitudes 
for all 19 of the dwarfs detected at Blanco and the NTT  are given in Table~\ref{tbl_measurements}. 


\section{Amalgamated sample and data}
\label{sec_sample}

The observations presented above and in \citet{fin10a} significantly expand the sample of dwarfs for which deep $K_s$-band photometry is available.  Thus, it is appropriate to re-examine the scaling relations elucidated earlier, especially to seek deeper insights into the why scatter is so much greater than expected from observational errors.

A sample of star-forming dwarfs was compiled from galaxies with extant $K_s$-band surface photometry to which a sech profile had been fitted.  In order to minimize scatter due to distance errors, the sample was restricted to objects for which the $I$ magnitude and $V-I$ colour of the tip of the red giant branch (TRGB) have been measured reliably.  A total of 66  
galaxies satisfied the criteria for analysis, and are listed in Table \ref{tbl_sample_distances}.  The galaxies IC~10 and ESO~245-05 were not included due to obvious problems with their photometry.   
In particular, IC~10 is a very large galaxy suffering from heavy extinction and severe crowding by foreground stars.  NGC~1560 was omitted because it is a late-type spiral.  

To establish homogeneous distances, the absolute magnitude of the TRGB was estimated from
\begin{eqnarray}
M_{I,TRGB} & = & (-3.935 \pm 0.028) \\
&  & + (0.217 \pm 0.020) \left[ (V-I)_{TRGB} - 1.6 \right]  \nonumber 
\end{eqnarray}
where $(V-I)_{TRGB}$ is the mean colour of the stars at the tip corrected for extinction and redshift \citep[see][]{riz07a}.  The zero-point was determined from a pairwise analysis of 
127 distances to 34 nearby galaxies  
derived from Cepheids, planetary nebulae, surface brightness fluctuations, and the TRGB (McCall, M. L., in preparation), and is anchored to the maser distance to NGC 4258 \citep[a distance modulus of 29.29:][]{her99a,gib00a,mac06a}.   
The uncertainty is that due to random errors only; it does not include the uncertainty in the distance to NGC 4258.  The rate of change of the absolute magnitude of the TRGB with colour was adopted from \citet{riz07a}.  

Apparent magnitudes and colours of TRGB stars were extracted from the literature.  In instances where colours were not recorded, they were estimated through inspection of colour-magnitude diagrams.  Where necessary, conversion of HST photometry to the Johnson-Cousins system was accomplished using the transformation equations of \citet{sir05a}.  Apparent magnitudes and colours were corrected for extinction and redshift (i.e., K-corrections) using the York Extinction Solver \citep{mcc04a}.  Optical depths were computed from $B-V$ colour excesses tabulated by \citet{sch98a}, and these were converted into extinctions and $V-I$ colour excesses assuming the spectral energy distribution of an M0 giant.  
The adopted reddening law was that of \citet{fit99a}, tuned to deliver a ratio of total-to-selective extinction of 
3.07  
for Vega \citep[see][]{mcc04a}.  Colour excesses, extinctions, corrected TRGB magnitudes and colours, and the resulting distance moduli for the 66 sample galaxies are recorded in Table~\ref{tbl_sample_distances}.  Adopted heliocentric velocities are given in Table~\ref{tbl_sample_observables}.  All K-corrections were less than $0.01 \, \rm mag$.  Uncertainties in distance moduli are estimated to be $0.10 \, \rm mag$  
typically.  

This paper extends work on absolute magnitudes of dwarfs into the realm of masses, motivated by the fact that dIs and BCDs retain a large fraction of their mass in gaseous form.  It is reasonable to expect that any correlation of absolute magnitude with dynamics will have scatter enhanced by variations in gas fractions, because the stellar component is often a minority of the visible mass (see Section~\ref{sec_mlvar}).   The determination of a baryonic mass requires that both the stellar and gaseous masses be constrained.

The mass of stars in each galaxy was judged to be most reliably signified by the luminosity in $K_s$, because the light from young stars is suppressed and the mass-to-light ratio is less sensitive to the star formation history than in bluer passbands \citep{joy88a,por04a,vad05a}.   The absolute magnitude of the diffuse stellar component was determined from the integrated magnitude of the sech function modeling the two-dimensional surface brightness profile.   The luminosity of any co-existing starburst was estimated by integrating the flux under the fitted Gaussian.  Corrections for extinction and redshift (i.e., K-corrections) were accomplished as for the TRGB, but using the Im spectral energy distribution of \citet{mcc04a}.  In computing luminosities from absolute magnitudes, the absolute magnitude of the Sun was adopted to be 
3.315   
in $K_s$ \citep{hol06a,fly06a}.   Note that no corrections for redshift dimming were applied.  All galaxies in the sample are at low redshift, and the entire range spanned by dimming corrections is only $0.01 \, \rm mag$.

Parameters describing the light distributions of the 66 sample galaxies are given in Table \ref{tbl_sample_observables}.  Listed are the corrected value $\mu_0$ of the central surface brightness in $\rm mag \, arcsec^{-2}$, the sech scale length $r_0$ converted to parsecs, the axis ratio $q$, the limiting radius of the surface photometry in units of $r_0$, and the source of the photometry.  The derived absolute magnitude $M_{Ks}$ of the sech component and the ratio of the luminosity of any burst relative to the luminosity of the sech component are given in Table~\ref{tbl_sample_derived}.  The adopted value of the extinction is included in Table~\ref{tbl_sample_distances}.  Generally, for galaxies observed on more than one occasion, parameters describing the fit to the deepest profile are presented.  However, parameters listed for ESO~381-18 and IC~4247  
come from averages of fits to two independent observations.

The mass of gas was determined from the integrated flux of HI at $21 \, \rm cm$.  Given the flux $F_{HI}$ in $\, \rm K \, km \, s^{-1}$, the mass $M_{gas}$ in solar units was computed from 
\begin{equation}
\label{eq_gasmass}
M_{gas} = k_{21} D_{Mpc}^2 F_{HI} / X  
\end{equation}
where $D_{Mpc}$ is the distance in Mpc, $X$ is the mass fraction of hydrogen, and 
$k_{21} = 2.356 \times 10^5 \, \rm M_\odot K^{-1} km^{-1} s$.
The value of $X$ was adopted to be 
0.735  
on the basis of measurements of the rate of change of the helium and metal fractions with the oxygen abundance in dwarfs \citep{izo10a}, presuming a mean oxygen abundance of 
8.25 for the current sample  
and a primordial helium mass fraction of 0.257  
\citep{izo10a}.  For galaxies for which multiple measurements of $F_{HI}$ were available, the single-dish measurement with the highest signal-to-noise ratio was normally adopted, unless there was evidence for confusion or peculiarities in the spectrum.  The adopted fluxes and sources are listed in Table~\ref{tbl_sample_observables}, and corresponding gas masses are given in Table~\ref{tbl_sample_derived}.

The width of the $21 \, \rm cm$ line at 20\% of the peak, $W_{20}$, was used to quantify internal motions.  The choice of $W_{20}$ over the width at 50\% of the peak was motivated by a desire to measure kinematics representative of the broadest possible body of gas.  

Earlier studies of the fundamental plane for dIs relied upon line widths with generally poor velocity resolution.  Most were taken from the Third Reference Catalogue of Bright Galaxies \citep{vau91a}.  For this paper, a comprehensive survey of the literature was made to pinpoint $21 \, \rm cm$ line profiles with the highest resolution and least noise.  Where 20\% line widths were not recorded, they were measured from the plotted profiles or, if justifiable, established mathematically from a fit used to determine the tabulated 50\% line width.  Because line profiles for the galaxies in this sample were very close to being Gaussian in shape, the apparent line width $W_{20}^{app}$ was corrected for instrumental broadening by subtracting in quadrature the width of a Gaussian instrumental profile as defined by the full-width at half-maximum $R$ \cite[see][]{ver01a}:
\begin{equation}
W_{20} = {W_{20}^{app} \over  (1 + z)} \sqrt{1 - {ln 5 \over ln 2} \left( R \over W_{20}^{app} \right)^2}  
\end{equation}
Here, $R$, $W_{20}^{app}$, and $W_{20}$ are in $\rm km \, s^{-1}$.
The factor of $1 + z$ corrects the width for redshift broadening.  The adopted values of the heliocentric velocity, $W_{20}^{app}$, $R$, and $\log W_{20}$, as well as the sources of the data, are given in Table~\ref{tbl_sample_observables}.

\section{Analysis}
\label{sec_analysis}

\subsection{Overview}
\label{sec_fitting}

Investigations below concentrate on elucidating how observed kinematics of dwarfs are tied to their scale and structure.   In the process, they lead to insights on what is driving gas motions, constraints on the mass-to-light ratio of stars, an evaluation of how close the galaxies are to being virialized, and the establishment of a method for determining reliable distances to unresolved objects.

\begin{figure*}[tbp]
\centering
\includegraphics[angle=0,width=13.5cm]{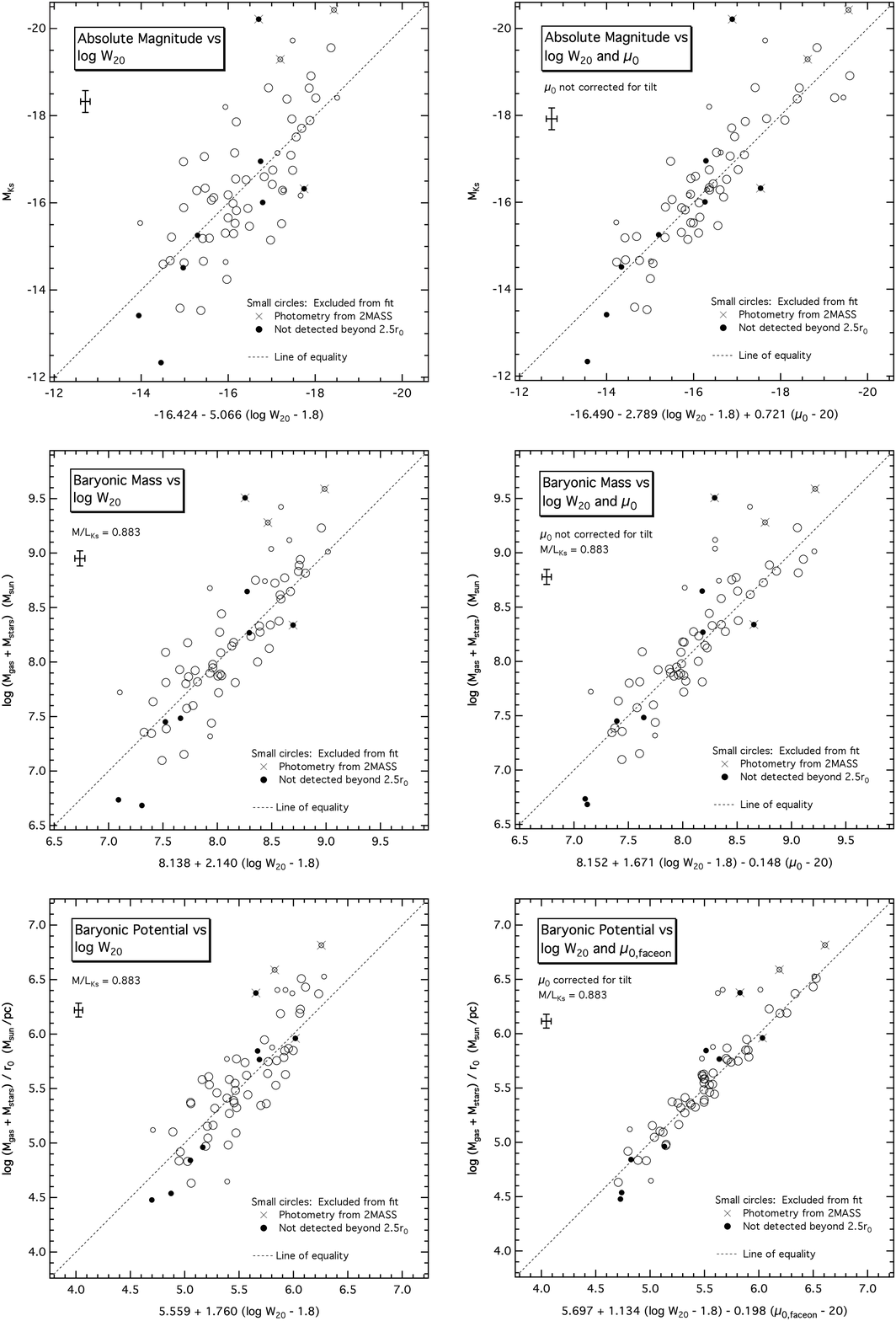}
\caption{
Correlations among intrinsic and observed properties of star-forming dwarfs.  In all panels, line widths have been corrected for resolution but not for tilt.  In magnitude units, the range of ordinates is the same for all panels, so vertical dispersions are directly comparable.  Typical uncertainties in abscissae and ordinates are depicted by an error cross in the upper left corner of each panel. {\it Top left:  } Absolute magnitude in $K_s$ versus the HI line width (the Tully-Fisher relation); {\it Top right:  } Absolute magnitude in $K_s$ versus the HI line width and the observed central surface brightness \citep[the {\it dI fundamental plane} of][with an updated fit]{vad05a}.  Correcting the surface brightnesses for tilt worsens the fit.; {\it Middle left:  }  Baryonic mass versus the HI line width, based upon a mass-to-light ratio in $K_s$ derived from the fit to the potential plane; {\it Middle right:  }  Baryonic mass versus the HI line width and the observed central surface brightness, based upon a mass-to-light ratio in $K_s$ derived from the fit to the potential plane.  Correcting the surface brightnesses for tilt does not improve the fit; {\it Bottom left:  } Baryonic potential versus the HI line width, based upon a mass-to-light ratio in $K_s$ derived from the fit to the potential plane; {\it Bottom right:  } Baryonic potential versus the HI line width and tilt-corrected central surface brightness (the {\it potential plane}), based upon a mass-to-light ratio in $K_s$ derived during the fitting.  In all panels, every galaxy in the sample is plotted, but only galaxies marked with large circles were fitted.  Excluded from the fits were four galaxies observed exclusively by 2MASS (crosses), six galaxies in addition to two 2MASS galaxies for which photometry did not extend beyond 2.5 sech scale lengths (solid circles),  one galaxy with an unusual morphology (small open circle), and seven extreme outliers identified while fitting the potential (small open circles).
}
\label{fig_six_correlations}
\end{figure*}

The most important correlations between intrinsic galaxy properties (absolute magnitude, mass, and potential) and distance-independent observables (HI line width, central surface brightness, and axis ratio) are displayed along with their fits in Figure~\ref{fig_six_correlations}.  The ordinates of the panels have been configured to span identical ranges in magnitude units, so that the apparent vertical dispersions about the fits are inter-comparable.  The displayed correlations are founded upon the properties of the sech component of the light profiles alone.  Derived global properties of the galaxies are summarized in Table~\ref{tbl_sample_derived}.  The relevance of the burst component and the distribution of the properties of bursting galaxies are discussed in Section~\ref{sec_bcds}.

The correlations were defined without correcting line widths for projection.  Even though it is known that some of the more massive dwarfs are rotating \citep[e.g.,][]{epi08a,swa09a}, there is considerable evidence that the kinetic energy of the gas in most of the galaxies in the sample is predominantly disordered \citep[][and this work]{vad05a}.  For example, line profiles tend to be Gaussian in shape, and the sensitivity of line widths to the ellipticity of isophotes is weak at best, implying that motions are close to isotropic (see Section~\ref{sec_potential}).

All galaxies in the sample are displayed in Figure~\ref{fig_six_correlations}, but only the 
48 galaxies   
marked by large open circles were included in fitting.  Photometry originating from 2MASS becomes suspect for galaxies whose surface brightnesses are as low as is typical of sample members \citep{kir08a}, and galaxies whose surface brightness profiles do not extend beyond 2.5 sech scale lengths  
tend to display deviant properties.  Consequently, the following galaxies were excluded from the fits:
\begin{itemize}
\item
four dwarfs observed by 2MASS only (Ho~II, NGC~3077, NGC~4214, and NGC~6822, marked by crosses)  
\item
in addition to two of the 2MASS galaxies (Ho~II and NGC~6822), six dwarfs for which photometry did not extend beyond $2.5 r_0$ (Cam~B, ESO444-84, KK98~230, KKH~86, Peg~DIG, and UGCA~92, marked by solid circles)  
\item
one object which displays a spiral-like morphology in HI and whose surface brightness profile is convex in the core (NGC~2915)
\item  
seven extreme deviants identified during the course of analysis (DDO~47, DDO~168, ESO215-09, ESO223-09, KK98~17, KK98~182, and UGC~3755), all of which are signified by small open circles   
\end{itemize}
The last seven galaxies were revealed by a large gap in the histogram of residuals for the fit to the potential versus line width and surface brightness.  They lie
$3.1 \sigma$ or more  
away from the fit, whereas the most extreme of the retained galaxies lie within 
$1.8 \sigma$  
of that fit.  

For maximum flexibility, fits were determined using the downhill simplex algorithm  \citep{nel65a,pre86a}.  For certain fits involving the stellar mass, it was postulated that any fundamental relationship is one for which there exists a mass-to-light ratio which minimizes the dispersion.  Stable solutions to the mass-to-light ratio proved to be possible by combining the simplex algorithm with a golden section search.  

Uncertainties in derived parameters were ascertained through Monte Carlo simulations.  The starting point for these simulations were estimates for typical errors in the observables.  The adopted uncertainties were 
$0.1 \, \rm mag$  
for distance moduli,  
$0.15 \, \rm mag$  
for $\mu_0$,  
5\% for $r_0$,  
10\% for $q$,  
$0.23 \, \rm mag$ for $m_{sech}$,  
5\% for $W_{20}$, and  
10\% for $F_{HI}$.   
In each analysis, 1000 random deviates of these observational quantities were computed, from which deviates for the derived quantities were computed and fitted.  Below, any quoted uncertainty in a fitted parameter is the average of the standard deviations of the resulting solutions on either side of the solution obtained from the reference fit.

\subsection{Absolute magnitude}
\label{absmag}

The upper left panel of Figure~\ref{fig_six_correlations} displays the Tully-Fisher relation for the dwarfs, albeit with no correction of line widths for tilt (as discussed above).  As expected, the dispersion is large.  The fitted relation is
\begin{eqnarray}
M_{Ks,sech} & = & (-16.424 \pm 0.040) \\
&  & - (5.066 \pm 0.207) (\log W_{20} - 1.8) \nonumber 
\end{eqnarray}
with the standard deviation being $0.95 \, \rm mag$.   
By comparison, the expected vertical scatter due to observational errors alone is only $0.27 \, \rm mag$   
(the quadrature sum of the uncertainties in the abscissa and ordinate).
Even if all motions were rotational, the dispersion in axis ratios is such that 
only about $0.5 \, \rm mag$ of the scatter would be attributable to projection.

The upper right panel of Figure~\ref{fig_six_correlations} displays the fundamental plane of \citet{vad05a}, with the fit updated.  For the sample here,
\begin{eqnarray}
M_{Ks, sech} & = & (-16.490 \pm 0.041) \\
&  & - (2.789 \pm 0.243) (\log W_{20} - 1.8) \nonumber \\
&  & + (0.721 \pm 0.043) (\mu_0 - 20) \nonumber 
\end{eqnarray}
where $\mu_0$ is the {\it observed} central surface brightness in $\rm mag \, arcsec^{-2}$ (i.e., without any correction for projection).
The standard deviation is $0.63 \, \rm mag$.   
The dispersion worsened when surface brightnesses were corrected for tilt (assuming an oblate spheroidal geometry: see Section~\ref{sec_potential}).  Efforts to improve the fit by attributing some of the motions to rotation, and correcting for tilt accordingly, met with failure.  The scatter is larger than suggested by \citet{vad08a}, probably because of changes to rejection criteria.  
Most importantly, as noted by \citet{vad08a}, it is much larger than the dispersion to be expected on the basis of observational errors alone, which is $0.28 \, \rm mag$.   

\subsection{Mass}

The unexplained scatter in absolute magnitudes motivated the development of a corresponding plane for the baryonic mass $M_{bary}$ which would accommodate the often significant but highly variable proportion of matter in gaseous form.   There was reason to be optimistic that a well-defined relationship might be found because the baryonic Tully-Fisher relation for rotationally-supported systems is so tight \citep{mcg00a,mcg05a,mcg11a}.  Most of the galaxies in the sample appear to be pressure-supported, with random motions being close to isotropic (see Section~\ref{sec_potential}), so on energy grounds it is reasonable to construct the baryonic Tully-Fisher relation by substituting radial velocity dispersions for circular velocities \citep{mcg10a,wol10a}.

The computation of baryonic masses required the adoption of a mass-to-light ratio for the stars, since
\begin{equation}
M_{bary} = \Upsilon L_{Ks} + M_{gas}  
\end{equation}
Here, $L_{Ks}$ is the luminosity in $K_s$ and $\Upsilon = M_{stars} / L_{Ks}$ is the mass-to-light ratio of the stars in $K_s$.  Attention was restricted to the sech component of the light distribution.  The contribution of a burst component to the mass was assumed to be negligible.  This approximation is justified in Section~\ref{sec_bcds}, where the consequences of accommodating the light of a burst are discussed.

The best estimate of $\Upsilon$ was gained from analysis of the gravitational potential (see Section~\ref{sec_potential}), which yielded a solution of $0.88 \pm 0.20$.   
To establish the most credible relationship between the baryonic mass and observables, then, the mass-to-light ratio was fixed at $0.88$.  
The corresponding stellar and baryonic masses are summarized in Table~\ref{tbl_sample_derived}.  

Surprisingly, the baryonic Tully-Fisher relation proved to be as highly dispersed as the fundamental plane.  It is displayed in the middle left panel of Figure~\ref{fig_six_correlations}.  The fitted relation is given by
\begin{eqnarray}
\label{eq_baryons_nomu0}
\log M_{bary} & = & (8.138 \pm 0.012) \\
&  & + (2.140 \pm 0.064) (\log W_{20} - 1.8) \nonumber  
\end{eqnarray}
The dispersion is
$0.24 \, \rm dex$ ($0.61 \, \rm mag$),  
versus the expected value of
$0.08 \, \rm dex$  
based upon observational uncertainties alone. 

An improvement to the baryonic Tully-Fisher relation 
was realized by introducing surface brightness as a second parameter.  The resulting baryonic plane, which is displayed in the middle right panel of Figure~\ref{fig_six_correlations}, is described by
\begin{eqnarray}
\label{eq_baryons}
\log M_{bary} & = & (8.152 \pm 0.012) \\
&  & + (1.671 \pm 0.074) (\log W_{20} - 1.8) \nonumber \\
&  & + (-0.148 \pm 0.013) (\mu_0 - 20) \nonumber 
\end{eqnarray}
(for $\Upsilon = 0.88$).
The standard deviation is $0.20 \, \rm dex$ ($0.49 \, \rm mag$).  
Although lower than found for the Tully-Fisher relation, it is still much higher than the expected value of $0.08 \, \rm dex$  
based upon observational uncertainties.   
The fit did not improve when the surface brightnesses were corrected for projection (see Section~\ref{sec_potential}).  Allowing $\Upsilon$ to be free, the dispersion dropped somewhat to $0.17 \, \rm dex$.   
However, the solution for $\Upsilon$ was $0.15 \pm 0.05$,   
which is unreasonably low compared to expectations from population syntheses \citep{por04a}.  

\subsection{Potential}
\label{sec_potential}

Thirty-two galaxies in the sample have line profiles which have been observed with a resolution of $2 \, \rm km \, s^{-1}$ or less.  Profiles are close to being Gaussian in shape, which suggests that the dynamics of dIs may be simple.  As a starting point, it is reasonable to posit that the systems are close to being virialized.  Virialization requires that
\begin{equation}
2T + \Omega = 0  
\end{equation}
where $T$ is the kinetic energy and $\Omega$ is the potential energy.
If the line width is predominantly controlled by gravity, and if the potential defined by the baryonic mass scales with the potential setting the line width (which in large part must be controlled by the amount of dark matter), then one might surmise that
\begin{equation}
\label{eq_pot_def}
P \equiv M_{bary} / r_0 \propto \left( W_{20} \right)^2  
\end{equation}
Henceforth, $P$ will be referred to as the ``baryonic potential''.

The baryonic potential (the stellar component of which being defined by the mass of the sech component) is plotted as a function of line width in the lower left panel of Figure~\ref{fig_six_correlations}.  The relationship is given by
\begin{eqnarray}
\label{eq_pot_nomu}
\log P & =& (5.559 \pm 0.011) \\
&  & + (1.760 \pm 0.058) (\log W_{20} - 1.8) \nonumber  
\end{eqnarray}
The standard deviation of the fit is 
$0.24 \, \rm dex$ ($0.59  \, \rm mag$),  
which, surprisingly, is comparable to that for the fits to the baryonic mass.  
However, the introduction of surface brightness as a second parameter reduced the dispersion drastically.  With the mass-to-light ratio fixed at 
0.88  
(see below), the following relationship was found:
\begin{eqnarray}
\label{eq_pot_noq}
\log P & =& (5.578 \pm 0.011) \\
&  & + (1.101 \pm 0.065) (\log W_{20} - 1.8) \nonumber \\
&  & + (-0.208 \pm 0.012) (\mu_0 - 20)  \nonumber 
\end{eqnarray}
The standard deviation is only 
$0.12 \, \rm dex$ ($0.29 \, \rm mag$).   
The dispersion did not change significantly when the mass-to-light ratio was allowed to vary.

\begin{figure*}tbp]
\centering
\includegraphics[angle=0,width=9cm]{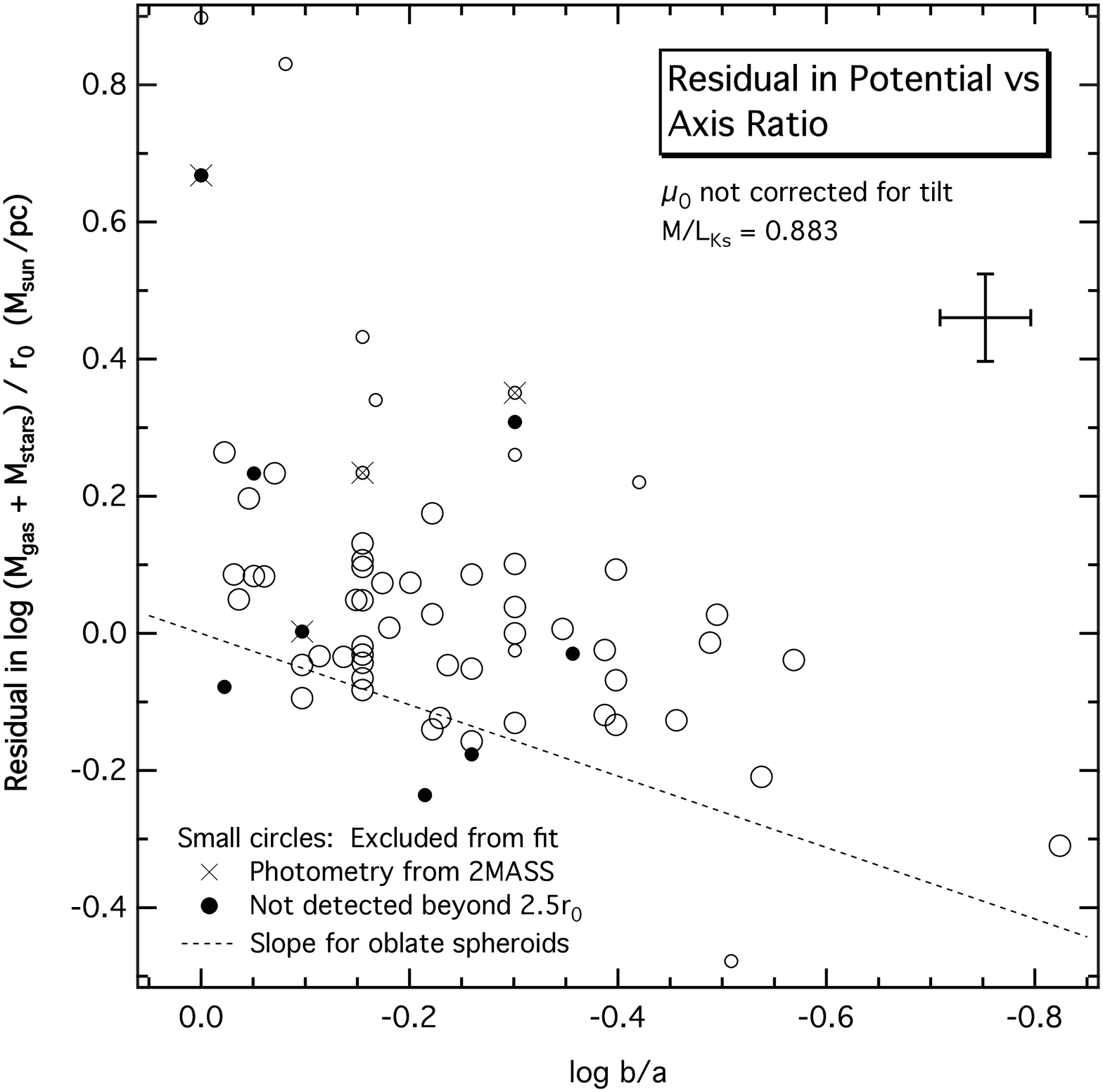}
\caption{
Residuals in the potential as a function of the axis ratio $q = b/a$.  The dashed line displays the slope of the expected relationship for oblate spheroids (the y-intercept having been fixed at zero).
} 
\label{fig_potential_qresid}
\end{figure*}

Figure~\ref{fig_potential_qresid} shows the residuals in the fit to the potential as a function of the logarithm of the axis ratio $q$.  
Residuals become more negative as galaxies flatten.  To a significant extent, this is likely to be a consequence of the effect of projection on surface brightnesses.  For an oblate spheroid, the surface brightness varies with $q$ as
\begin{equation}
\label{eq_mu0_project}
\mu_0 = \mu_{0}^{i=0} + 2.5 \log q  
\end{equation}
where $\mu_{0}^{i=0}$ is the surface brightness that would be measured if the view were face-on (inclination $i$ equal to zero).  The dashed line in Figure~\ref{fig_potential_qresid} displays the rate at which residuals in the potential should vary with $q$ if surface brightnesses are affected by projection in the way expected for oblate spheroids.  The observed trend is very close to that predicted.   Thus, a further refinement to the fit to the potential was possible by correcting surface brightnesses to a common viewing angle (face-on).  

It was unclear how line widths dominated by random motions might vary with tilt.  For the purpose of investigation, any sensitivity to tilt was approximated as a power law in $q$.  Then, to convert measurements to face-on,
\begin{equation}
W_{20}^{i=0} = q^\gamma W_{20}  
\end{equation}
where $\gamma$ is a constant.
It was expected that a relationship between the potential, line width, and surface brightness which is free of projection effects would have the form
\begin{eqnarray}
\log P & = & a  + b \log q^\gamma W_{20} + c (\mu_0 - 2.5 \log q) \\
& =  & a + b \log W_{20} + c \mu_0 + (\gamma b - 2.5 c) \log q  
\end{eqnarray}
where $a$, $b$, and $c$ are constants.  By introducing $\log q$ as a third variable, it was possible to constrain $\gamma$.

With $\Upsilon$ fixed at  
0.88,   
and with the geometry approximated to be oblate spheroidal,
the solution for $\gamma$ was $-0.10$.   
The sign is opposite to what would be expected if flattening is a consequence of anisotropic motions, be they ordered or disordered.  Also, $| \gamma |$ is small, suggesting that motions are close to being isotropic.  It was concluded that the trend in the residuals of the potential with the axis ratio should be attributed primarily to variations in surface brightness expected for oblate spheroids viewed at different angles. 

In the end, given how weakly line widths appeared to depend on tilt and uncertainty about precisely how they did, only surface brightnesses were corrected for projection.  The relationship between the potential, apparent line width, and the surface brightness corrected to face-on (via Equation~\ref{eq_mu0_project}) is displayed in the lower right panel of Figure~\ref{fig_six_correlations}.  With the mass-to-light ratio free to vary, the fit was given by
\begin{eqnarray}
\label{eq_pot_vs_mufo}
\log P & = & (5.697 \pm 0.065) \\
&  & + (1.134 \pm 0.080) (\log W_{20} - 1.8) \nonumber \\
&  & + (-0.198 \pm 0.018) (\mu_0^{i=0} - 20)  \nonumber 
\end{eqnarray}
The corresponding solution for $\Upsilon$ was
$0.883 \pm 0.199$. 
Resulting values of $\log M_{stars}$, $\log M_{bary}$, and $\log P$ are listed in Table~\ref{tbl_sample_derived}.  Uncertainties in the coefficients are higher than for Equation~\ref{eq_pot_noq} because of the freedom in the mass-to-light ratio.  The standard deviation is only 
$0.096 \, \rm dex$ ($0.24 \, \rm mag$).   
The vertical dispersion expected from random observational errors is $0.08 \, \rm dex$.  
Thus, a structural relationship has been identified for dwarfs for which observational errors overwhelm the cosmic dispersion.  This relationship can be regarded as a {\it more} fundamental plane for dwarfs, and henceforth will be referred to as the {\it potential plane}.

Although the solution for the mass-to-light ratio is twice as high as measured for the disk of the Milky Way from vertical kinematics of stars \citep{por09a}, it is nevertheless compatible with syntheses of exponential disks spanning a range of possible star formation histories \citep{por04a}.  It should be noted, though, that the estimate for $\Upsilon$ is in part a dynamical estimate of the mass-to-light ratio, since it is tied to line widths.  The tightness of the correlation lends credence to the postulate that the mass and size of dark matter haloes scale straightforwardly with the mass and distribution of luminous matter.

To check the sensitivity of the fit to the selection of galaxies, random subsets of the original galaxy sample were formed by removing 10\% of the objects (5 galaxies).  A total of 100 subsets were constructed and fitted.  The mean values of the free parameters agreed extremely well with those determined from the entire sample.  Most important, the mean value of $\Upsilon$ was $0.895 \pm 0.095$,  
which is almost identical to the value derived by fitting the whole sample.  Because observational errors are predominant in setting the dispersion, it is not believed that the solution for the mass-to-light ratio suffers from biases which plague fits to less fundamental relations.

\subsection{Blue compact dwarfs and turbulence}
\label{sec_bcds}

Despite much research, the relationship between dIs and BCDs is not 
clear yet.  Structurally, they appear to be similar, because the near-infrared light profile of a BCD can be modeled well by superimposing a Gaussian starburst upon a sech function \citep{vad06a}.  Thus, it is reasonable to consider any star-forming dwarf to be a blue compact dwarf if its light profile in $K_s$ displays an excess of light in the core over what is expected for a pure sech law.  That is the definition adopted here.

\begin{figure*}[tbp]
\centering
\includegraphics[angle=0,width=18cm]{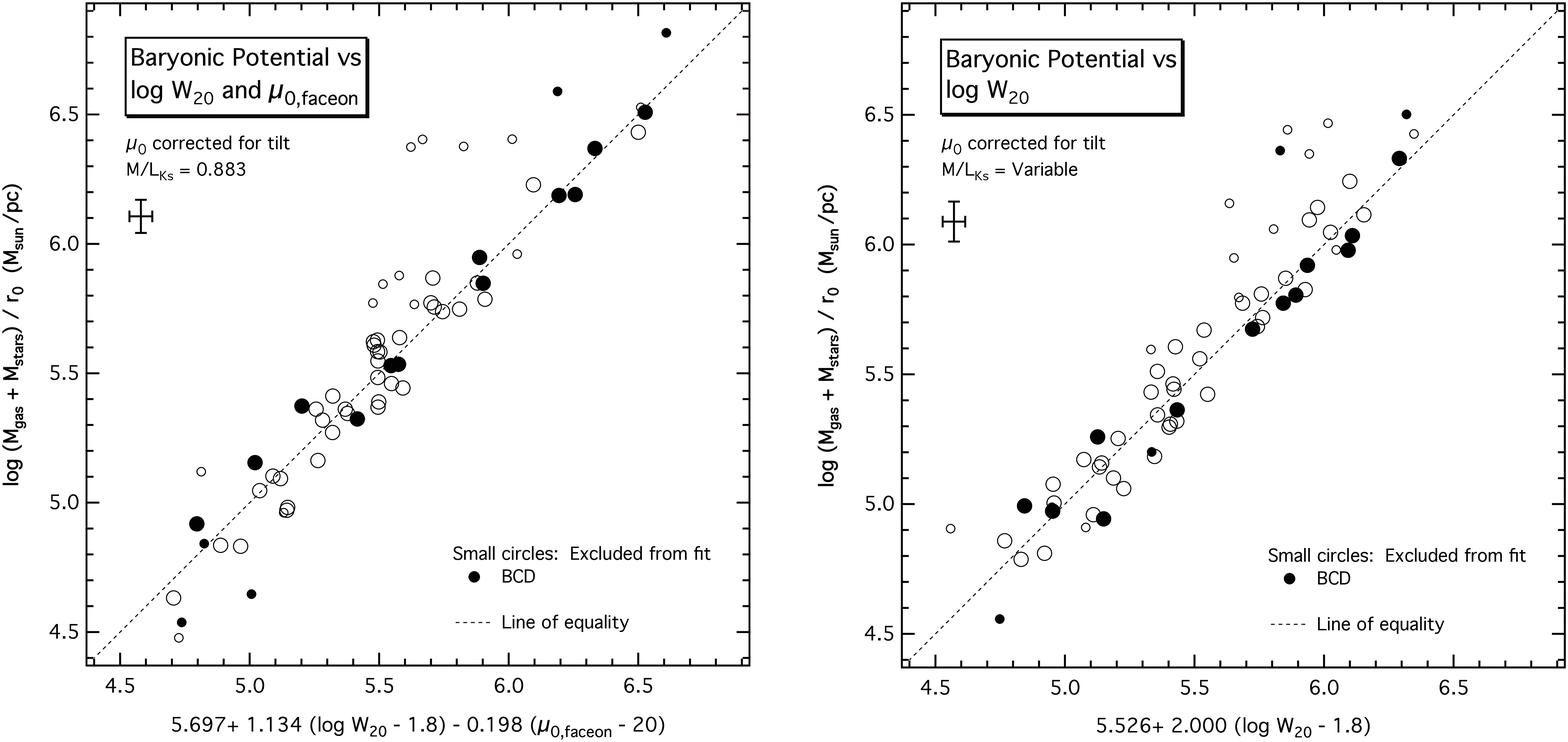}
\caption{
Two renditions of the potential for dwarfs.  {\it Left}:  $M_{stars} / L_{Ks}$ fixed.  {\it Right}: $M_{stars} / L_{Ks}$ variable.  Individualized mass-to-light ratios for the plot on the right were estimated from the amount by which the central surface brightness of the sech component deviated from the mean for a given line width under the condition that the galaxies are virialized (see Figure~\ref{fig_mu0ref}).  In both panels, dwarf irregulars (dIs) are marked with open circles, and blue compact dwarfs (BCDs) are flagged by solid circles.  No matter how the mass-to-light ratios are computed, BCDs and dIs populate the diagrams in the same way.
}
\label{fig_bcds}
\end{figure*}

The left panel of Figure~\ref{fig_bcds} re-displays the correlation of the potential with line width and face-on central surface brightness (Equation~\ref{eq_pot_vs_mufo}).  Those galaxies with an excess of light in the core, i.e., the BCDs,  are marked with solid circles, and normal dIs are marked with open circles.  Starbursts span the entire range of galaxies in the sample.  Fits to the pure sech dwarfs and the BCDs separately were consistent within errors, proving that there is no segregation.   It appears that gravity, not turbulence, is the predominant determinant of gas kinematics in most of the star-forming dwarfs in the sample.  The mere existence of strong correlations of the baryonic mass and the potential with line width lends support to this conclusion.  Also, the dependence of the potential on surface brightness is opposite to what would be expected if line widths were inflated by gas flows stemming from recent star formation.  For a gas-dominated system at least, any concomitant enhancement in surface brightness would be expected to be incorporated in such a way as to oppose the change in line width and thereby preserve the potential.  Instead, the sign of the coefficient of $\mu_0^{i=0}$ in Equation~\ref{eq_pot_vs_mufo} is such that brightening and broadening change the potential in the same way. 

To examine the influence of the Gaussian component on the fit to the potential, the mass-to-light ratio of stars in the burst relative to the mass-to-light ratio of stars in the sech was introduced as a free parameter.  Then, luminosities of both the burst and sech components were employed to compute baryonic masses.  The solution for the ratio of mass-to-light ratios was zero, meaning that the inclusion of a non-negligible burst mass degrades the correlation.  However, forcing the ratio of mass-to-light ratios to unity led to a fit which was still reasonable.  A larger sample of BCDs in which the burst light is a significant fraction of the total light will be required to constrain more reliably the appropriate mass-to-light ratio to use to accommodate the burst mass in the definition of the potential.

\section{Discussion}
\label{sec_discussion}

\subsection{Variable mass-to-light ratios and virialization}
\label{sec_mlvar}

Equation~\ref{eq_pot_vs_mufo} reveals that
\begin{equation}
\label{eq_pot_vs_w20i}
P \propto \left[ W_{20} \right] ^{1.13} \left[ I_0^{i=0} \right]^{0.49}  
\end{equation}
where $I_0^{i=0}$ is the face-on central surface brightness of the sech model in linear units.  Although the virial theorem motivated the quest for this relation, the exponent of $W_{20}$ is half what it should be.  The dependence on the surface brightness is puzzling, too.  Possibly, it is a reaction to variations in the mass-to-light ratio.  A dwarf which has undergone star formation more recently than is typical for galaxies of its kind in the sample may show enhancements in both surface brightness and luminosity relative to the norm for its potential.  If so, the mass-to-light ratio ought to be reduced by a factor which preserves the value of the potential.  In effect, such a reduction is happening through the dependence of $P$ on $I_0^{i=0}$, although it is complicated by the presence of gas.  It is even possible that, in not accounting for variability in $\Upsilon$, the true dependence of the potential on the line width is obscured.    

To individualize mass-to-light ratios, and thereby assess the impact on the relationship between the potential and internal motions, the premise was made that any deviation of the surface brightness from the mean at a given line width is a consequence of a different star formation history, and that the deviation is simultaneously incorporated in the luminosity of stars associated with the sech component.  Then, the mass-to-light ratio $\Upsilon$  relative to some reference value $\Upsilon_{ref}$ could be estimated from the face-on central surface brightness $\mu_0^{i=0}$ relative to an appropriate norm $\mu_{0,ref}^{i=0}$ as follows:
\begin{equation}
\label{eq_mlvar}
\log (\Upsilon / \Upsilon_{ref}) = 0.4 (\mu_0^{i=0} - \mu_{0,ref}^{i=0})  
\end{equation}

\begin{figure*}[tbp]
\centering
\includegraphics[angle=0,width=9cm]{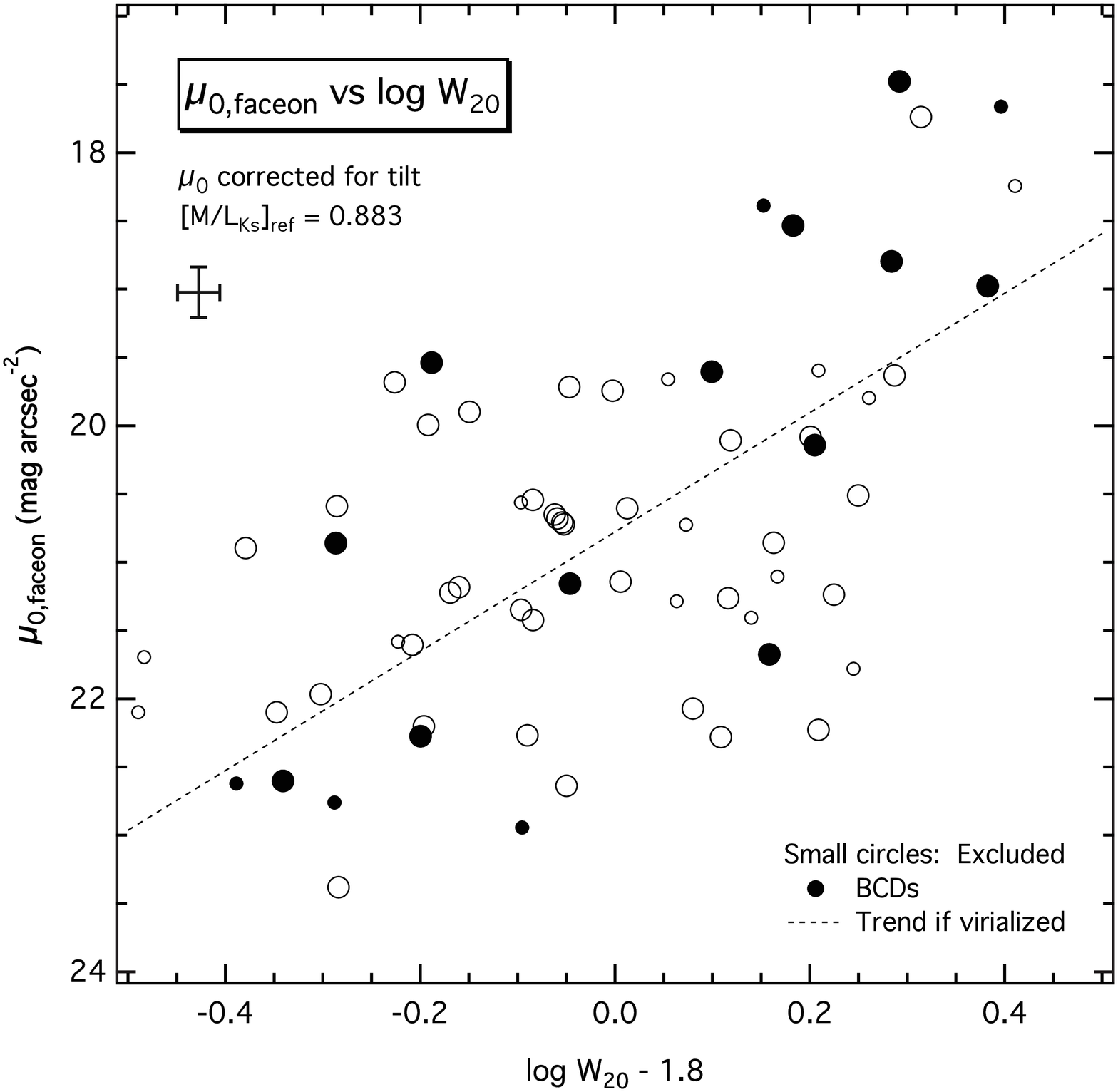}
\caption{
Tilt-corrected central surface brightness of the sech component as a function of the HI line width.   Dwarf irregulars (dIs) are marked with open circles, and blue compact dwarfs (BCDs) are flagged by solid circles.  The dashed line is the required relationship if the galaxies are virialized, based upon a reference mass-to-light ratio of 0.883.
}
\label{fig_mu0ref}
\end{figure*}

In principle, the mean surface brightness  $\mu_{0,ref}^{i=0}$ could be tied to global properties.  Figure~\ref{fig_mu0ref} reveals the correlation of $\mu_0^{i=0}$ with $\log W_{20}$.  There is a tendency for surface brightnesses to brighten with the line width, although the scatter is large.  The correlation must in part be responsible for weakening the $W_{20}$-dependence of the potential relative to what is expected for virialized systems.

If it is hypothesized that star-forming dwarfs really are virialized, it is possible to predict how rapidly $\mu_{0,ref}^{i=0}$ must vary with $\log W_{20}$, and thereby test compatibility with the observations.  To this end, it was approximated that $\mu_{0,ref}^{i=0}$ varies linearly with $\log W_{20}$.  Based upon the earlier fit to the potential with $\Upsilon$ held fixed, it was reasonable to approximate $\Upsilon_{ref}$ to be 0.883.  Noting that $\Upsilon$ for each dwarf is defined by Equation~\ref{eq_mlvar}, it was possible to solve for the coefficients of the relation between $\mu_{0,ref}^{i=0}$ and $\log W_{20}$ by minimizing the dispersion in the relationship between the potential and $\log W_{20}$.

The dashed line in Figure~\ref{fig_mu0ref} displays the solution for the correlation between $\mu_{0,ref}^{i=0}$ and $\log W_{20}$ which arose when the galaxies were required to be virialized.  It is defined by 
\begin{eqnarray}
\label{eq_mu0ref}
\mu_{0,ref}^{i=0} & = & (20.778 \pm 0.214)  \\
&  & + (-4.371 \pm 0.296) (\log W_{20} - 1.8)  \nonumber 
\end{eqnarray}
The data admit the possibility of such a trend.
The right panel of Figure~\ref{fig_bcds} shows the simultaneous fit to the potential, with baryonic masses now determined using individualized mass-to-light ratios computed from Equations~\ref{eq_mlvar} and \ref{eq_mu0ref} (with $\Upsilon_{ref}$ set to 0.883).  It is described by
\begin{equation}
\label{eq_pot_vs_w20_mlvar}
\log P = (5.526 \pm 0.041) + 2 (\log W_{20} - 1.8)  
\end{equation} 
Of course, the coefficient in front of $\log W_{20}$ was forced by the requirement that the systems be virialized.  The standard deviation about the fit is 
$0.105 \, \rm dex$ ($0.26 \, \rm mag$),   
which is only a little worse than that for the potential plane (Equation~\ref{eq_pot_vs_mufo}).  

\begin{figure*}[tbp]
\centering
\includegraphics[angle=0,width=9cm]{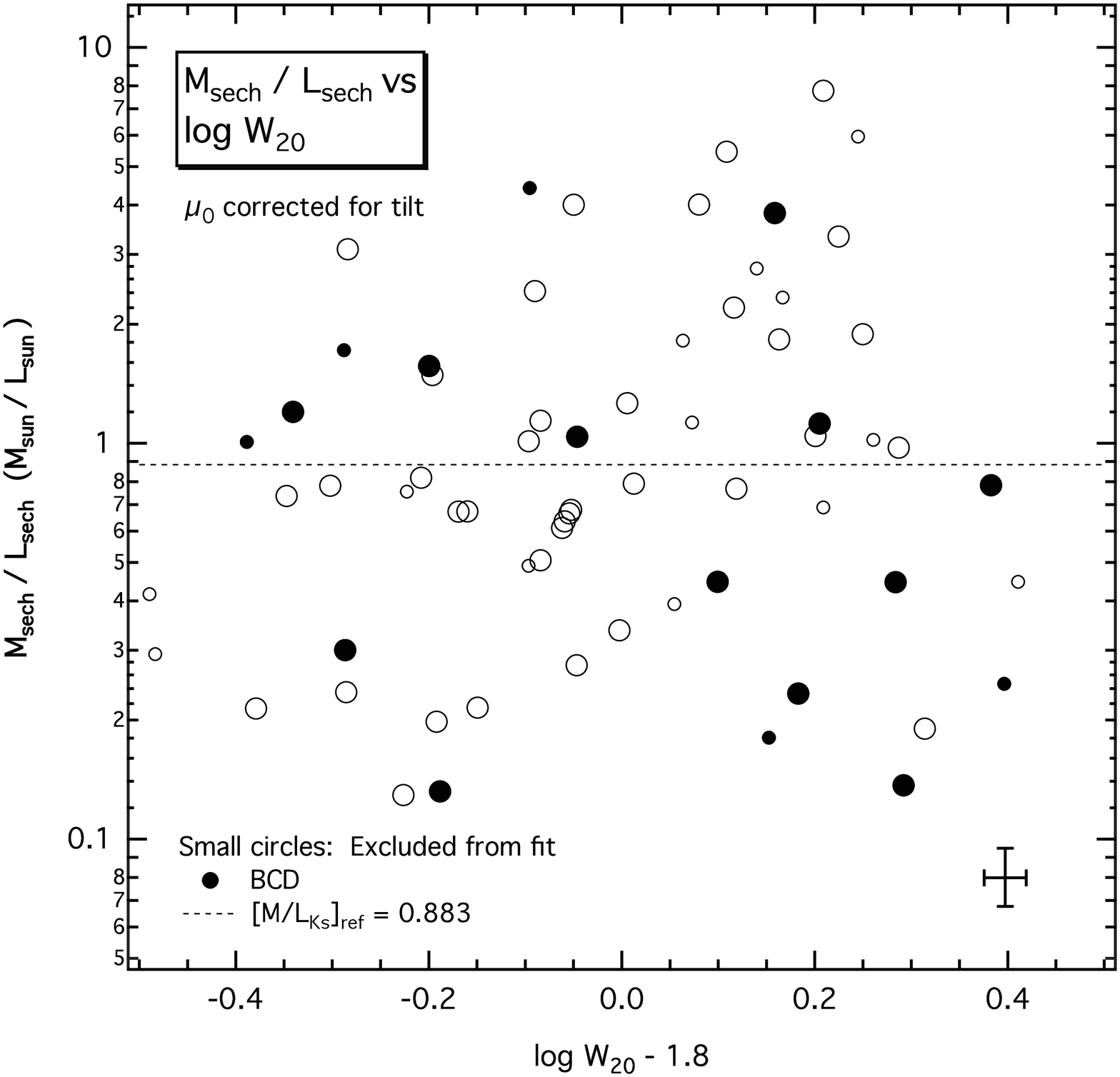}
\caption{
Mass-to-light ratio of the sech component as a function of line width.  Dwarf irregulars (dIs) are marked with open circles, and blue compact dwarfs (BCDs) are flagged by solid circles. Individualized mass-to-light ratios were determined from the deviations in surface brightness from the norms expected for virialized systems.  The horizontal dashed line is the value of the mass-to-light ratio which was computed for the entire sample in establishing the potential plane of Figure~\ref{fig_six_correlations}.
}
\label{fig_mlvar}
\end{figure*}

Resulting mass-to-light ratios are listed in Table~\ref{tbl_sample_derived} and displayed as a function of $\log W_{20}$ in Figure~\ref{fig_mlvar}.  The range of variation is fairly large, in many cases beyond what is reasonable to expect in $K_s$ for typical star formation histories \citep{por04a}.  Observational errors are in part to blame, since mass-to-light ratios vary to compensate for errors in surface brightness.

\begin{figure*}[tbp]
\centering
\includegraphics[angle=0,width=18cm]{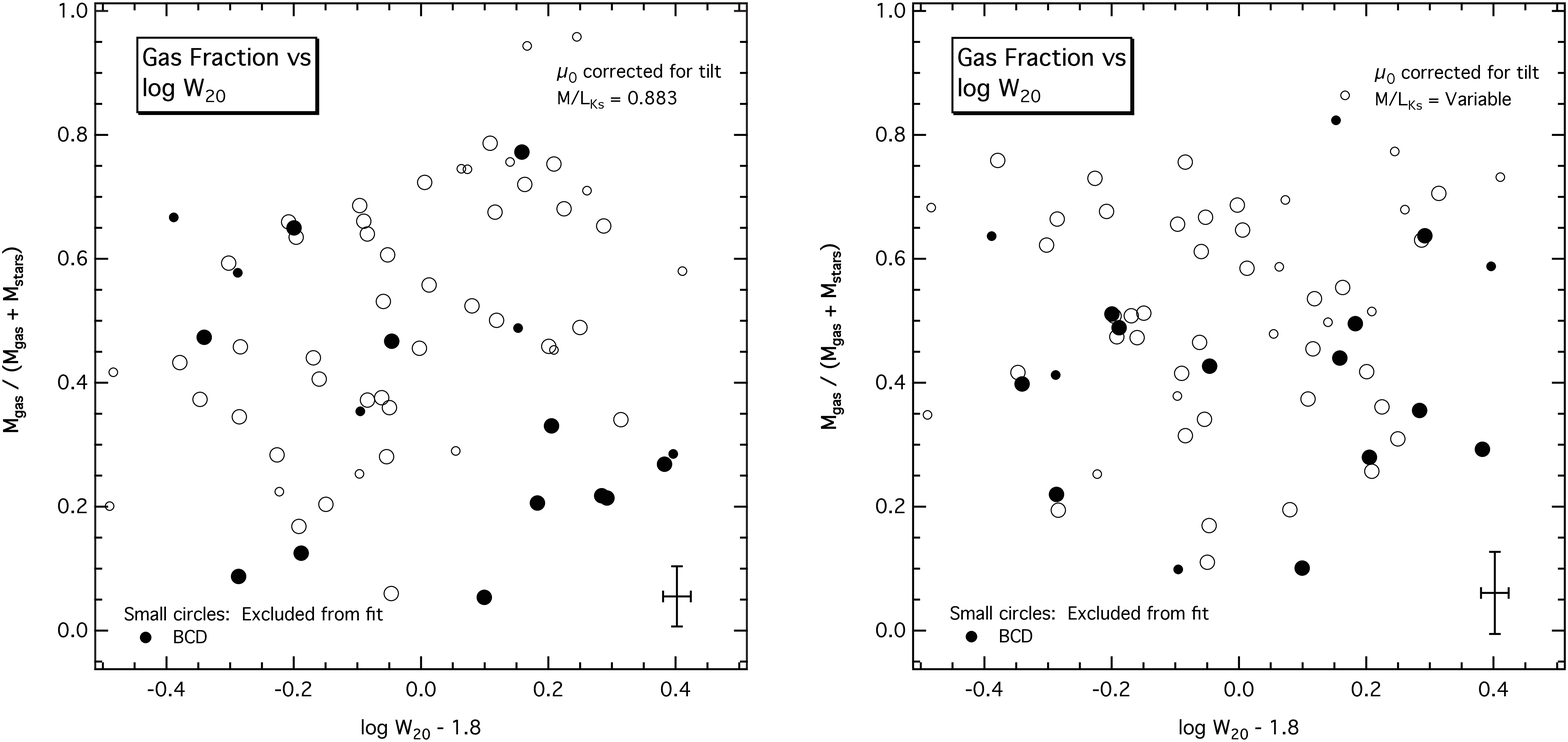}
\caption{
Gas fractions as a function of line width. {\it Left}:  $M_{stars} / L_{Ks} $ fixed.  {\it Right}: $M_{stars} / L_{Ks}$ variable.  In both panels, dwarf irregulars (dIs) are marked with open circles, and blue compact dwarfs (BCDs) are flagged by solid circles.  With mass-to-light ratios adjusted for surface brightness deviations, gas fractions for many galaxies change significantly.
}
\label{fig_gasfrac}
\end{figure*}

Figure~\ref{fig_gasfrac} compares the gas fractions computed from a fixed value of the mass-to-light ratio ($\Upsilon  = 0.88$) with the gas fractions derived when the mass-to-light ratio is allowed to vary as above. The spread in gas fractions at large $W_{20}$ is reduced when variable mass-to-light ratios are employed.  For many star-bursting dwarfs, the gas fraction rises.  This is because surface brightnesses for these galaxies are unusually high, leading to downward adjustments to mass-to-light ratios and consequent reductions in the stellar masses.  Gas fractions for many dwarf irregulars also change significantly, rising for the least massive galaxies and declining for the most massive.

\subsection{Distances}

The discovery of tight correlations between the gravitational potential and distance-independent observables opens up a way of mapping the spatial distribution of dwarf galaxies on large scales.  Locally, dwarfs are more dispersed than giants, so the relationship offers an avenue for exploring the distribution of matter in mass-poor regions of the universe.  A huge advantage over the Tully-Fisher relation for spirals is that no restriction need be placed on tilt.   In fact, the method for correcting surface brightness for the viewing angle does not even require that the tilt be evaluated (a complicated problem owing to the possible variability of the intrinsic axis ratio).  

For any galaxy, the value of $P \equiv M_{bary} / r_0$  in $\rm M_\odot \, pc^{-1}$ can be estimated from observations of the $21 \, \rm cm$ line width $\log W_{20}$, the central surface brightness $\mu_0$, and the axis ratio $q$ using Equation~\ref{eq_pot_vs_mufo} or \ref{eq_pot_vs_w20_mlvar}.   This provides an avenue to determining the distance.  Define $m_{sech}$ to be the apparent magnitude of the sech component of the light distribution corrected for extinction and redshift, $r_{0,arcsec}$  the sech scale length in arc seconds, $\Upsilon$ the mass-to-light ratio in solar units for the stars constituting the sech component (for the passband defining $m_{sech}$), $F_{21}$ the $21 \, \rm cm$ line flux in $\rm K \, km \, s^{-1}$, $X$ the fraction of the gas mass which is hydrogen, and $k_{21}$ the factor required to convert the HI line flux to hydrogen mass.  The distance modulus $D_{mod}$ is given by
\begin{eqnarray}
\label{eq_distance}
D_{mod} / 5 & = & \log P  + \log r_{0,arcsec} \\
&  & - \log \left[ \Upsilon 10^{-0.4 (m_{sech} - M_\odot - 12.5)}
+ 10^{-5} k_{21} F_{21} / X \right] \nonumber \\
&  & - \log (0.648 / \pi) \nonumber  
\end{eqnarray}
where $M_\odot$ is the absolute magnitude of the Sun 
(3.315 in $K_s$)  
and $k_{21}$ has the value specified for Equation~\ref{eq_gasmass}.  For $K_s$, the mass-to-light ratio is either fixed at 
0.883  
if Equation~\ref{eq_pot_vs_mufo} is used to estimate $P$, or computed from Equations \ref{eq_mlvar} and \ref{eq_mu0ref} if Equation~\ref{eq_pot_vs_w20_mlvar} is employed to get $P$ (with $\Upsilon_{ref}$ set to 0.883).  

\begin{figure*}[tbp]
\centering
\includegraphics[angle=0,width=9cm]{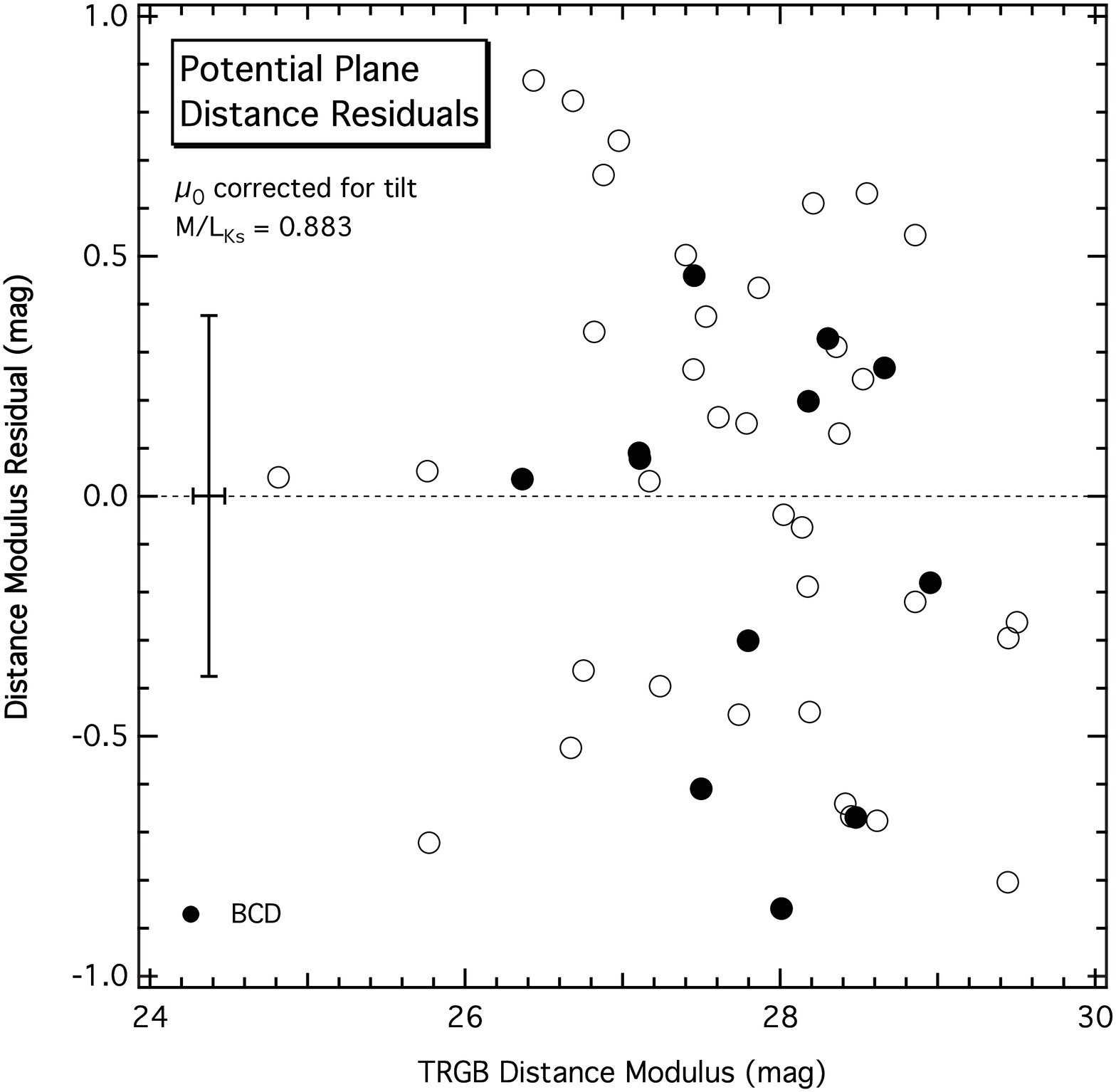}
\caption{
Difference between the potential plane distance modulus and TRGB distance modulus as a function of the TRGB distance modulus for dwarfs defining the potential plane.  The horizontal dashed line marks perfect agreement.  Dwarf irregulars (dIs) are marked with open circles, and blue compact dwarfs (BCDs) are flagged by solid circles. The vertical component of the error bar shows the mean uncertainty expected for potential plane distances from observational errors if the cosmic scatter about the potential plane is zero.}
\label{fig_dmod_resid}
\end{figure*}

Figure~\ref{fig_dmod_resid} shows for the galaxies defining the potential plane how the distances derived from Equation~\ref{eq_distance} compare with those derived from the TRGB.
Typical uncertainties in observational quantities were given in Section~\ref{sec_fitting}.  Because observational errors are responsible for the bulk of the dispersion in the potential, it is reasonable to examine the accuracy with which a distance can be determined considering observational errors alone.  Based upon the uncertainties in $W_{20}$, $\mu_0$, and $q$, $\log P$ can be estimated to an accuracy of about 
$0.04 \, \rm dex$  
(Equation~\ref{eq_pot_vs_mufo}) if the cosmic dispersion is smaller.  Accounting for the errors in $m_{sech}$, $r_{0,arcsec}$ and $F_{21}$, then the uncertainty in the derived distance modulus comes out to be 
$0.38 \, \rm mag$   
for 
$\Upsilon = 0.88$.    

The most important contributors to the uncertainty are the errors in $m_{sech}$ and $\mu_0$.  In the sample studied here, these errors were quite large due to limitations in the field of view, which restricted the number of 2MASS stars available to calibrate the photometry.  Significant improvements are possible using more modern detectors with wider fields of view.  If the uncertainty in $m_{sech}$ can be reduced to 
$0.10 \, \rm mag$   
and the uncertainty in 
$\mu_0$ to $0.05 \, \rm mag$,  
then the error in the distance modulus will come down to 
$0.26 \, \rm mag$.    
A further reduction is possible with deeper photometry, which would enable improvement in the accuracy with which $r_0$ and $q$ are measured.  Overall, it appears that the potential plane for dwarfs  offers a means to determine distances to dwarfs as good as the Tully-Fisher relation yields for spirals.


\section{Conclusions}
\label{sec_conclusions}
Deep imaging in $K_s$ has been presented for 19 star-forming dwarf galaxies.  Structural properties were measured by fitting a sech function to surface brightness profiles, and additionally a Gaussian for those objects displaying evidence for a central starburst (interpreted thereby as being blue compact dwarfs).  Results for these galaxies were combined with photometry for others published previously to examine how kinematics, as conveyed by HI line widths, are connected to global properties.

Most dwarfs in the sample displayed HI line profiles close to Gaussian in shape.  Also, in optimizing relationships among global properties and kinematics, no strong tie between apparent line widths and isophotal axis ratios was evident.  Thus, in the majority of sample galaxies, most of the kinetic energy of the gas appeared to be disordered, and internal motions appeared to be close to isotropic.  Consequently, it proved to be possible to establish relationships without correcting line widths for tilt.

It was confirmed that much of the scatter in the Tully-Fisher relation for dwarfs is correlated with surface brightness.  The ``fundamental plane'' defined by the correlation of absolute magnitude with line width and surface brightness still displayed a dispersion which exceeded what was expected from observational errors alone.  Conjecturing that some of the scatter might be a consequence of not accommodating the highly variable gaseous component, the baryonic Tully-Fisher relation was constructed.  It proved to be as dispersed as the fundamental plane, although some improvement was possible by adding surface brightness as a second parameter.  

Motivated by the possibility that the galaxies may be virialized, the correlation between the potential and the line width was examined.   The potential defined by the ratio of the baryonic mass to the sech scale length was hypothesized to be proportional to the potential setting internal motions, but the mass-to-light ratio required to compute the stellar component of the mass was left as a free parameter.  The derived relationship between the potential and the line width displayed large residuals correlated with surface brightness.  When surface brightness variations were accommodated, an extremely tight relationship between the potential, line width, and surface brightness resulted.  Remaining residuals were found to correlate with tilt, with the bulk of the trend explainable by variations in surface brightness arising from viewing oblate spheroids at different angles.  Once surface brightnesses were corrected for tilt, the remaining dispersion of this {\it more} fundamental plane, referred to as the {\it potential plane}, could be almost entirely attributable to observational errors.   The solution for the mean mass-to-light ratio for the stars (i.e., the sech component of the light distribution) was $0.88 \pm 0.20$.  BCDs lay precisely in the plane described by dIs.   The placement of BCDs, the strength of the correlation, and the direction of the sensitivity to surface brightness all point to gravity, not turbulence, being primarily responsible for determining gas motions in star-forming dwarfs.  The potential plane suggests a strong linkage between the mass and distribution of luminous matter and the mass and scale of dark matter haloes.  It also offers a new avenue for determining the distances to unresolvable star-forming dwarfs which may be as good as the Tully-Fisher relation for spirals.  

The potential plane described a potential varying as 
$W_{20}^{1.13}$,  
i.e., less steeply than expected for virialized systems (for which the exponent should be 2).  However, the dependence of the potential on surface brightness was such that surface brightness might be considered to be acting as a proxy for variations in the mass-to-light ratio of stars.  To explore this possibility, it was hypothesized that star-forming dwarfs are in fact virialized, and the required dependence of mean surface brightness on line width was derived.  Deviations in surface brightness from the norm for a particular line width were used to adjust mass-to-light ratios to compensate for differing star formation histories (and observational errors).  Residuals in the correlation of the modified potential with line width were only slightly degraded with respect to those for the potential plane.  Computed mass-to-light ratios covered a range greater than expected from theoretical expectations, but some of the variation was attributable to the uncertainties in surface brightnesses.  Further studies of star formation rates would be productive in evaluating whether the predicted range of mass-to-light ratios is justified.  Work is in progress to evaluate implications for chemical evolution.


\begin{acknowledgements}
MLM is grateful to M. De Robertis for advice on testing the robustness of the fit to the potential plane, and to the Natural Sciences and Engineering Council of Canada for its continuing support. 
OV, EUS, MA, FPN and ABD acknowledge the Chilean TACs for the time allocation at ESO (Run ID: 081.B-0386(A)) and CTIO (2008A-0913 and 2008B-0909). 
This research has made use of the NASA/IPAC Extragalactic Database (NED) which is operated by the Jet Propulsion Laboratory, California Institute of Technology, under contract with the National Aeronautics and Space Administration. 
We acknowledge the usage of the HyperLeda database (http://leda.univ-lyon1.fr - Paturel et al. 2003). 
Our work used IRAF, a software package distributed by the National Optical Astronomy Observatory, which is operated by the Association of Universities for Research in Astronomy (AURA) under cooperative agreement with the National Science Foundation. 
This research has made use of SAOImage DS9, developed by the Smithsonian Astrophysical Observatory. 
\end{acknowledgements}





\clearpage

\begin{table*}
\caption{Log of Observations} 
\label{tbl_obs_log}
\centering
\begin{tabular}{lccccc}
\hline
\hline
\noalign{\smallskip}
Galaxy & RA (2000) & DEC (2000) & Observatory & Date (UT) & Exposure (s) \\
\hline
\noalign{\smallskip}

 ESO 294-10      & 00:26:33.4 & $-$41:51:19 & NTT    & Aug 16, 2008 & 2340 \\ 
 ESO 540-30      & 00:49:20.9 & $-$18:04:32 & NTT    & Aug 16, 2008 & 2340 \\ 
 UGC 685         & 01:07:22.4 & +16:41:04 & NTT    & Aug 16, 2008 & 2280 \\
 ESO 245-05      & 01:45:03.7 & $-$43:35:53 & BLANCO & Aug 12, 2008 & 2400 \\
 NGC 1311        & 03:20:06.9 & $-$52:11:08 & NTT    & Aug 16, 2008 & 1920 \\
 UGC 2684        & 03:20:23.7 & +17:17:45 & NTT    & Aug 16, 2008 & 2340 \\
 ESO 121-20      & 06:15:54.2 & $-$57:43:32 & BLANCO & Mar 13, 2008 & 1020 \\
 ESO 121-20      &    ...     &     ...   & BLANCO & Aug 12, 2008 & 2640 \\
 ESO 059-01      & 07:31:18.2 & $-$68:11:17 & BLANCO & Mar 13, 2008 & 1020 \\
 NGC 2915        & 09:26:11.5 & $-$76:37:35 & BLANCO & Mar 13, 2008 & 1020 \\
 Antlia Dwarf    & 10:04:04.1 & $-$27:19:52 & BLANCO & Mar 14, 2008 & 1980 \\
 ESO 215-09      & 10:57:29.9 & $-$48:10:43 & BLANCO & Mar 13, 2008 & 1980 \\
 ESO 320-14      & 11:37:53.2 & $-$39:13:13 & BLANCO & Mar 13, 2008 & 2340 \\
 ESO 379-07      & 11:54:43.5 & $-$33:33:36 & BLANCO & Mar 13, 2008 & 1980 \\
 ESO 443-09      & 12:54:54.0 & $-$28:20:27 & BLANCO & Mar 13, 2008 & 1980 \\
 ESO 381-18      & 12:44:42.4 & $-$35:58:00 & BLANCO & Mar 14, 2008 & 1020 \\
 ESO 321-14      & 12:13:49.6 & $-$38:13:53 & BLANCO & Mar 13, 2008 & 2220 \\
 KK98~182        & 13:05:02.1 & $-$40:04:58 & BLANCO & Mar 14, 2008 & 1980 \\
 KK98~195        & 13:21:08.2 & $-$31:31:45 & NTT    & Aug 16, 2008 & 1500 \\
 IC 4247         & 13:26:44.4 & $-$30:21:45 & BLANCO & Mar 14, 2008 & 1020 \\
 ESO 444-78      & 13:36:31.1 & $-$29:14:06 & BLANCO & Mar 13, 2008 & 1020 \\
 HIPASS J1337$-$39 & 13:37:25.3 & $-$39:53:48 & BLANCO & Mar 13, 2008 & 1020 \\
 HIPASS J1337$-$39 &     ...    &     ...   & BLANCO & Aug 12, 2008 & 1260 \\
 IC 4316         & 13:40:18.4 & $-$28:53:32 & BLANCO & Mar 12, 2008 & 1020 \\
 HIPASS J1348$-$37 & 13:48:47.0 & $-$37:58:29 & BLANCO & Mar 14, 2008 & 1980 \\
 HIPASS J1351$-$47 & 13:51:12.0 & $-$46:58:13 & BLANCO & Mar 12, 2008 & 1020 \\ 
 HIPASS J1351$-$47 &    ...     &     ...   & BLANCO & Mar 14, 2008 & 1980 \\ 
 NGC 5408        & 14:03:20.9 & $-$41:22:40 & BLANCO & Aug 11, 2008 & 1200 \\
 PGC 51659       & 14:28:03.7 & $-$46:18:06 & BLANCO & Mar 12, 2008 & 2580 \\
 ESO 137-18      & 16:20:58.4 & $-$60:29:28 & BLANCO & Aug 12, 2008 & 1260 \\
 Sag DIG         & 19:29:59.0 & $-$17:40:41 & BLANCO & Aug 12, 2008 & 2460 \\
 DDO 210         & 20:46:51.8 & $-$12:50:53 & BLANCO & Aug 12, 2008 & 1260 \\
 DDO 210         &    ...     &     ...   & NTT    & Aug 16, 2008 & 2340 \\
 ESO 468-020     & 22:40:43.9 & $-$30:48:00 & NTT    & Aug 16, 2008 & 2280 \\
 ESO 149-03      & 23:52:02.8 & $-$52:34:40 & NTT    & Aug 16, 2008 & 2040 \\
\hline
\end{tabular}
\end{table*}


\begin{table*}
\caption{
Measurements for Detected Galaxies}
\label{tbl_measurements}
\centering
\begin{tabular}{lcccccc}
\hline
\hline
\noalign{\smallskip}
Galaxy & 
$\mu_{0K}$ & 
$r_{0K}$ & 
$m_{SK}$ & 
$m_{TK}$ & 
PA & 
$e$ 
\\

& 
$\rm (mag \, arcsec^{-2})$ & 
(arcsec) & 
(mag) & 
(mag) & 
(deg) & 
\\

(1) & 
(2) & 
(3) & 
(4) & 
(5) & 
(6) & 
(7)  
\\
\hline
\noalign{\smallskip}

 ESO 294-10      & 20.53 &  12.2 & 12.66 & 12.87 &  ~~\phantom{+}0 & 0.2 \\
 UGC 685         & 19.74 &  14.8 & 11.59 & 11.72 & $-$55 & 0.3 \\
 ESO 245-05      & 21.09 &  30.5 & 11.57 & 11.65 & $-$60 & 0.4  \\
 NGC 1311        & 19.15 &  22.4 & 10.72 & 10.55 & +40 & 0.6  \\
 UGC 2684        & 21.17 &  11.9 & 14.10 & 14.15 & $-$65 & 0.6 \\
 ESO 121-20      & 20.48 &  13.4 & 12.58 & 12.66 & +50 & 0.3 \\
 ESO 059-01      & 19.89 &  18.3 & 11.16 & 11.23 &  ~~\phantom{+}0 & 0.2  \\
 NGC 2915        & 17.59 &  17.6 &  ~~9.46 &  ~~9.26 & $-$40 & 0.5  \\
 ESO 320-14      & 20.31 &  10.6 & 12.92 & 12.81 & +90 & 0.3  \\
 ESO 381-18      & 20.17 &   ~~8.9 & 13.16 & 13.15 & +80 & 0.3 \\
 ESO 321-14      & 20.52 &  18.3 & 12.85 & 12.87 & +30 & 0.7 \\
 KK98~182        & 20.76 &  ~~8.7 & 13.81 & 14.06 & +90 & 0.3 \\
 IC 4247         & 18.77 &  14.5 & 11.61 & 11.64 & $-$30 & 0.7 \\
 ESO 444-78      & 20.14 &  13.3 & 12.61 & 12.52 & +30 & 0.5 \\
 IC 4316         & 20.32 &  26.8 & 11.08 & 11.09 & +45 & 0.4  \\
 NGC 5408        & 18.90 &  19.2 & 10.59 & 10.69 & +75 & 0.5 \\
 ESO 137-18      & 18.31 &  24.3 &  ~~9.48 &  ~~9.40 & +30 & 0.5  \\
 ESO 468-20      & 21.29 &  17.4 & 12.95 & 13.07 & +40 & 0.4  \\
 ESO 149-03      & 20.07 &  11.3 & 13.11 & 13.22 & $-$30 & 0.6 \\

\hline
\end{tabular}
\tablefoot{
(1) Name of galaxy;
(2) Central surface brightness of sech model in $K_s$;  
(3) Scale length of sech model in $K_s$;
(4) Apparent magnitude of sech model in $K_s$;
(5) Apparent isophotal magnitude in $K_s$; 
(6) Position angle of major axis (from N through E); 
(7) Ellipticity of isophotes.
}
\end{table*}


\longtab{3}{
\begin{longtable}{lccccccccc}
\caption{\label{tbl_sample_distances}
Sample for analysis:  Extinctions and Distances}
\\
\hline
\hline
\noalign{\smallskip}

Galaxy &
$E(B-V)$	&
$A_V^{gal}$ &
$A_I^{gal}$ &
$A_{Ks}^{gal}$	&
$I_{TRGB}$ &
$(V - I)_{TRGB}$ &
$M_{I,TRGB}$	&
$D_{mod}$ &
Source
\\ 

& 
(mag) & 
(mag) & 
(mag) & 
(mag) & 
(mag) & 
(mag) &
(mag) &
(mag) &
\\
 
(1) & 
(2) & 
(3) & 
(4) & 
(5) & 
(6) &
(7) &
(8) &
(9) &
(10)
\\

\hline

\endfirsthead
\caption{continued.}\\
\hline
\hline

\noalign{\smallskip}

Galaxy &
$E(B-V)$	&
$A_V^{gal}$ &
$A_I^{gal}$ &
$A_{Ks}^{gal}$	&
$I_{TRGB}$ &
$(V - I)_{TRGB}$ &
$M_{I,TRGB}$	&
$D_{mod}$ &
Source
\\ 

& 
(mag) & 
(mag) & 
(mag) & 
(mag) & 
(mag) & 
(mag) &
(mag) &
(mag) &
\\
 
(1) & 
(2) & 
(3) & 
(4) & 
(5) & 
(6) &
(7) &
(8) &
(9) &
(10)
\\

\hline
\noalign{\smallskip}

\endhead
\hline
\endfoot

\noalign{\smallskip}

\object{Cam~B}	 &	0.219	&	0.695	&	0.394	&	0.082	&	23.60	&	1.65	&	 $-$3.92	&	27.52	&	1	\\
\object{CGCG~087-33}	&	0.032	&	0.102	&	0.058	&	0.012	&	25.47	&	1.15	&	 $-$4.03	 &	29.50	&	2	\\
\object{DDO~006}	&	0.017	&	0.054	&	0.031	&	0.006	&	23.57	&	1.48	&	 $-$3.96	 &	27.53	&	3	 \\
\object{DDO~047}	&	0.033	&	0.105	&	0.059	&	0.012	&	25.48	&	1.13	&	 $-$4.04	 &	29.52	&	2	\\
\object{DDO~099}	&	0.026	&	0.083	&	0.047	&	0.010	&	23.11	&	1.29	&	$-$4.00 	&	27.11	&	4	\\
\object{DDO~167}	&	0.010	&	0.032	&	0.018	&	0.004	&	24.06	&	1.49	&	 $-$3.96	 &	28.02	&	5	\\
\object{DDO~168}	&	0.015	&	0.048	&	0.027	&	0.006	&	24.13	&	1.40	&	 $-$3.98 	&	28.11	&	5	\\
\object{DDO~181}	&	0.006	&	0.019	&	0.011	&	0.002	&	23.46	&	1.35	&	 $-$3.99	 &	27.45	&	4	\\
\object{DDO~187}	&	0.023	&	0.073	&	0.041	&	0.009	&	22.69	&	1.34	&	 $-$3.99	 &	26.68	&	4	\\
\object{DDO~190}	&	0.012	&	0.039	&	0.022	&	0.005	&	23.19	&	1.38	&	 $-$3.98	 &	27.17	&	4	\\
\object{DDO~226}	&	0.015	&	0.049	&	0.028	&	0.006	&	24.41	&	1.47	&	 $-$3.96	 &	28.37	&	3	\\
\object{ESO~059-01}	&	0.147	&	0.467	&	0.265	&	0.055	&	24.26	&	1.54	&	 $-$3.95	 &	28.21	&	6	 \\
\object{ESO~121-20}	&	0.042	&	0.132	&	0.075	&	0.016	&	24.86	&	1.33	&	 $-$3.99	 &	28.86	&	6	\\
\object{ESO~137-18}	&	0.245	&	0.777	&	0.441	&	0.092	&	25.01	&	1.55	&	 $-$3.95 	&	28.95	&	7	\\
\object{ESO~215-09}	&	0.221	&	0.700	&	0.397	&	0.083	&	24.58	&	1.43	&	 $-$3.97	 &	28.55	&	7	\\
\object{ESO~223-09}	&	0.260	&	0.824	&	0.467	&	0.097	&	25.04	&	1.59	&	 $-$3.94	 &	28.98	&	7	\\
\object{ESO~269-58}	&	0.109	&	0.345	&	0.196	&	0.041	&	23.86	&	1.61	&	 $-$3.93	 &	27.80	&	8	\\
\object{ESO~320-14}	&	0.143	&	0.454	&	0.257	&	0.053	&	24.89	&	1.45	&	 $-$3.97	 &	28.86	&	7	\\
\object{ESO~321-14}	&	0.094	&	0.299	&	0.170	&	0.035	&	23.51	&	1.34	&	 $-$3.99	 &	27.50	&	4	\\
\object{ESO~324-24}	&	0.113	&	0.358	&	0.203	&	0.042	&	23.81	&	1.43	&	 $-$3.97	 &	27.79	&	9	\\
\object{ESO~325-11}	&	0.088	&	0.279	&	0.158	&	0.033	&	23.62	&	1.36	&	 $-$3.99	 &	27.61	&	9	\\
\object{ESO~349-31}	&	0.012	&	0.038	&	0.022	&	0.004	&	23.48	&	1.43	&	 $-$3.97	 &	27.45	&	6	\\
\object{ESO~379-07}	&	0.074	&	0.236	&	0.134	&	0.028	&	24.54	&	1.59	&	 $-$3.94	 &	28.48	&	9	\\
\object{ESO~381-18}	&	0.063	&	0.199	&	0.113	&	0.023	&	24.58	&	1.46	&	 $-$3.97	 &	28.55	&	7	\\
\object{ESO~381-20}	&	0.066	&	0.208	&	0.118	&	0.024	&	24.64	&	1.42	&	 $-$3.98	 &	28.61	&	7	\\
\object{ESO~384-16}	&	0.074	&	0.235	&	0.133	&	0.028	&	24.23	&	1.55	&	 $-$3.95	 &	28.18	&	7	\\
\object{ESO~444-78}	&	0.053	&	0.167	&	0.095	&	0.020	&	24.55	&	1.42	&	 $-$3.97	 &	28.53	&	7	\\
\object{ESO~444-84}	&	0.069	&	0.218	&	0.123	&	0.026	&	24.27	&	1.29	&	 $-$4.00	 &	28.28	&	9	\\
\object{ESO~461-36}	&	0.303	&	0.962	&	0.546	&	0.114	&	25.45	&	1.32	&	 $-$4.00	 &	29.45	&	6 \\
\object{GR~8}	&	0.026	&	0.082	&	0.046	&	0.010	&	22.63	&	2.21	&	 $-$3.80 	&	26.43	&	2	\\
\object{Ho~II}	&	0.032	&	0.101	&	0.057	&	0.012	&	23.61	&	1.41	&	 $-$3.98	 &	27.59	&	4	\\
\object{IC~3104}	&	0.410	&	1.301	&	0.739	&	0.154	&	22.75	&	1.29	&	 $-$4.00	 &	26.75	&	10	\\
\object{IC~4247}	&	0.065	&	0.205	&	0.116	&	0.024	&	24.43	&	1.39	&	 $-$3.98	 &	28.41	&	7	\\
\object{IC~4316}	&	0.055	&	0.173	&	0.098	&	0.020	&	24.10	&	1.72	&	 $-$3.91	 &	28.01	&	9	\\
\object{IC~4662}	&	0.070	&	0.222	&	0.126	&	0.026	&	22.89	&	1.64	&	 $-$3.93	 &	26.82	&	6	\\
\object{IC~5152}	&	0.025	&	0.079	&	0.045	&	0.009	&	22.42	&	1.58	&	 $-$3.94	 &	26.36	&	2	\\
\object{KK98~17}	&	0.055	&	0.173	&	0.098	&	0.020	&	24.41	&	1.07	&	 $-$4.05	 &	28.46	&	2	\\
\object{KK98~182}	&	0.102	&	0.325	&	0.184	&	0.038	&	24.77	&	1.29	&	 $-$4.00	 &	28.77	&	7	\\
\object{KK98~200}	&	0.069	&	0.219	&	0.124	&	0.026	&	24.20	&	1.60	&	 $-$3.93	 &	28.14	&	9	\\
\object{KK98~230}	&	0.014	&	0.045	&	0.025	&	0.005	&	22.47	&	1.31	&	 $-$4.00	 &	26.47	&	4	\\
\object{KKH~086}	&	0.027	&	0.085	&	0.048	&	0.010	&	23.09	&	1.32	&	 $-$4.00	 &	27.08	&	4	\\
\object{KKH~098}	&	0.123	&	0.389	&	0.221	&	0.046	&	22.92	&	1.49	&	 $-$3.96 	&	26.88	&	10	\\
\object{Mrk~178}	&	0.019	&	0.060	&	0.034	&	0.007	&	23.90	&	1.49	&	 $-$3.96	 &	27.86	&	5	\\
\object{NGC~1311}	&	0.022	&	0.068	&	0.039	&	0.008	&	24.63	&	1.15	&	 $-$4.03	 &	28.66	&	2	\\
\object{NGC~1569}	&	0.694	&	2.206	&	1.254	&	0.262	&	23.12	&	1.36	&	 $-$3.99	 &	27.10	&	11	\\
\object{NGC~2915}	&	0.275	&	0.872	&	0.495	&	0.103	&	23.87	&	1.76	&	 $-$3.90	 &	27.77	&	12	\\
\object{NGC~3077}	&	0.067	&	0.212	&	0.120	&	0.025	&	23.93	&	1.77	&	 $-$3.90	 &	27.82	&	4	\\
\object{NGC~3738}	&	0.010	&	0.033	&	0.019	&	0.004	&	24.40	&	1.75	&	 $-$3.90	 &	28.30	&	5	\\
\object{NGC~4163}	&	0.020	&	0.063	&	0.036	&	0.007	&	23.27	&	1.44	&	 $-$3.97	 &	27.24	&	4	\\
\object{NGC~4214}	&	0.022	&	0.069	&	0.039	&	0.008	&	23.43	&	1.47	&	 $-$3.96 	&	27.39	&	4	\\
\object{NGC~5408}	&	0.068	&	0.216	&	0.123	&	0.025	&	24.36	&	0.90	&	 $-$4.09	 &	28.45	&	9	\\
\object{NGC~6822}	&	0.231	&	0.732	&	0.415	&	0.086	&	19.39	&	1.58	&	 $-$3.94	 &	23.32	&	13	\\
\object{Peg~DIG}	&	0.068	&	0.216	&	0.122	&	0.025	&	20.75	&	1.52	&	 $-$3.95	 &	24.70	&	14	\\
\object{Sex~A}	&	0.045	&	0.141	&	0.080	&	0.017	&	21.76	&	1.23	&	 $-$4.02 	&	25.77	&	4	\\
\object{Sex~B}	&	0.031	&	0.099	&	0.056	&	0.012	&	21.77	&	1.35	&	 $-$3.99 	&	25.76	&	4	\\
\object{UGC~0685}	&	0.057	&	0.182	&	0.103	&	0.021	&	24.32	&	1.12	&	 $-$4.04	 &	28.36	&	2	\\
\object{UGC~3755}	&	0.088	&	0.280	&	0.159	&	0.033	&	25.31	&	1.08	&	 $-$4.05	 &	29.36	&	2	\\
\object{UGC~4115}	&	0.028	&	0.090	&	0.051	&	0.011	&	25.39	&	1.04	&	 $-$4.06	 &	29.44	&	2	\\
\object{UGC~4483}	&	0.034	&	0.108	&	0.061	&	0.013	&	23.72	&	1.23	&	 $-$4.02	 &	27.74	&	4	\\
\object{UGC~6456}	&	0.037	&	0.119	&	0.067	&	0.014	&	24.20	&	1.38	&	 $-$3.98	 &	28.19	&	15	\\
\object{UGC~7605}	&	0.015	&	0.046	&	0.026	&	0.005	&	24.18	&	1.34	&	 $-$3.99	 &	28.17	&	5	\\
\object{UGC~8508}	&	0.015	&	0.048	&	0.027	&	0.006	&	22.99	&	1.38	&	 $-$3.98 	&	26.98	&	10	\\
\object{UGC~8833}	&	0.012	&	0.037	&	0.021	&	0.004	&	23.41	&	1.36	&	 $-$3.99	 &	27.40	&	4	\\
\object{UGCA~092}	&	0.785	&	2.496	&	1.420	&	0.296	&	23.46	&	1.18	&	 $-$4.03	 &	27.49	&	6	\\
\object{UGCA~438}	&	0.015	&	0.046	&	0.026	&	0.005	&	22.69	&	1.37	&	 $-$3.98 	&	26.67	&	4	\\
\object{WLM}	&	0.038	&	0.120	&	0.068	&	0.014	&	20.85	&	1.47	&	 $-$3.96	 &	24.82	&	16	\\

\hline

\end{longtable}

\tablefoot{
(1) Name of galaxy;  
(2) Galactic colour excess from \citet{sch98a};
(3) Galactic extinction of M0 III star in V;
(4) Galactic extinction of M0 III star in I; 
(5) Galactic extinction of dwarf irregular galaxy in $K_s$; 
(6) $I$ magnitude of stars at tip of red giant branch, corrected for extinction and redshift;
(7) $(V-I)$ colour of stars at tip of red giant branch, corrected for extinction and redshift;
(8) Absolute magnitude of TRGB stars in $I$;
(9) Distance modulus;
(10) Source of colour-magnitude diagram.
}

\tablebib{
(1) \citet{kar03d}; 
(2) \citet{tul06a};
(3) \citet{kar03c};
(4) \citet{dal09a};
(5) \citet{kar03a};
(6) \citet{kar06a};
(7) \citet{kar07a};
(8) \citet{dav07a};
(9) \citet{kar02a};
(10) \citet{kar02b};
(11) \citet{gro08a};
(12) \citet{kar03b};
(13) \citet{gal96a};
(14) \citet{mcc05a};
(15) \citet{men02a};
(16) \citet{riz07a}.
}

}




\longtab{4}{
\begin{longtable}{lccccccccccc}
\caption{\label{tbl_sample_observables}
Sample for Analysis:  Radio and Optical Parameters
}
\\

\hline
\hline
\noalign{\smallskip}

Galaxy &
$V_\odot$ &
$F_{HI}$ &	
$W_{20}^{app}$ &	
$R$	&	
$\log W_{20}$ &	
HI &
$\mu_0$ &	
$r_{0,pc}$	&	
$q$ &	
$r_{lim} / r_0$ &	
Photometry
\\

& 
($\rm km \, s^{-1}$) & 
($\rm Jy \, km \, s^{-1}$) & 
($\rm km \, s^{-1}$) & 
($\rm km \, s^{-1}$) & 
($\rm km \, s^{-1}$) & 
Source &
($\rm mag \, arcsec^{-2}$) &
(pc) &
&
&
Source
\\
 
(1) & 
(2) & 
(3) & 
(4) & 
(5) & 
(6) &
(7) &
(8) &
(9) &
(10) &
(11) &
(12)
\\

\hline

\endfirsthead
\caption{continued.}\\
\hline
\hline

\noalign{\smallskip}

Galaxy &
$V_\odot$ &
$F_{HI}$ &	
$W_{20}^{app}$ &	
$R$	&	
$\log W_{20}$ &	
HI &
$\mu_0$ &	
$r_{0,pc}$	&	
$q$ &	
$r_{lim} / r_0$ &	
Photometry
\\

& 
($\rm km \, s^{-1}$) & 
($\rm Jy \, km \, s^{-1}$) & 
($\rm km \, s^{-1}$) & 
($\rm km \, s^{-1}$) & 
($\rm km \, s^{-1}$) & 
Source &
($\rm mag \, arcsec^{-2}$) &
(pc) &
&
&
Source
\\
 
(1) & 
(2) & 
(3) & 
(4) & 
(5) & 
(6) &
(7) &
(8) &
(9) &
(10) &
(11) &
(12)
\\

\hline
\noalign{\smallskip}

\endhead
\hline
\endfoot

\noalign{\smallskip}

Cam~B	&	~~78 	&	~~5.0	&	32.6	&	~~1.65	&	1.512	&	H1, H2, H3	&	21.87	&	407	&	0.44	&	2.40	&	P1	\\
CGCG~087-33	&	279	 &	~~2.6	&	55.8	&	1.4	&	1.746	&	H1	&	19.74	&	532	&	0.41	&	4.13	&	P2	\\
DDO~006	&	295	 &	~~3.4	&	32.9	&	1.4	&	1.516	&	H1	&	22.04	&	570	&	0.29	&	2.53	&	P3	\\
DDO~047	&	272	 &	61.4	&	111.0~~	&	~~2.06	&	2.045	&	H4 &	21.78	&	555	&	1.00	&	4.20	&	P4	\\
DDO~099	&	243 	&	46.0	&	91.0	&	~~2.06	&	1.958	&	H4	&	21.30	&	393	&	0.71	&	2.67	&	P1	\\
DDO~167	&	163	&	~~4.6	&	40.2	&	1.4	&	1.604	&	H1	&	21.65	&	337	&	0.60	&	3.47	&	P4	\\
DDO~168	&	191	&	74.4	&	88.0	&	~~8.24	&	1.940	&	H5, H6	&	20.36	&	733	&	0.38	&	3.96	&	P1	\\
DDO~181	&	202	&	11.4	&	56.8	&	1.4	&	1.754	&	H1	&	20.77	&	356	&	0.70	&	2.87	&	P1	\\
DDO~187	&	153	&	12.0	&	51.4	&	1.4	&	1.710	&	H1	&	21.71	&	287	&	0.60	&	3.68	&	P4	\\
DDO~190	&	151	&	27.1	&	64.0	&	~~8.24	&	1.797	&	H5, H6	&	19.62	&	257	&	0.89	&	4.21	&	P1	\\
DDO~226	&	359	&	~~6.1	&	56.4	&	~~1.65	&	1.750	&	H7, H3	&	20.58	&	982	&	0.15	&	3.05	&	P3	\\
ESO~059-01	&	530	&	17.7	&	104.0~~	&	18~~~~	&	2.001	&	H8	&	19.84	&	389	&	0.80	&	4.92	&	P5	\\
ESO~121-20	&	577	&	14.1	&	96.0	&	18~~~~ 	&	1.963	&	H8	&	20.47	&	384	&	0.70	&	4.48	&	P5	\\
ESO~137-18	&	605	&	37.4	&	155.0~~	&	18~~~~	&	2.183	&	H8	&	18.22	&	727	&	0.50	&	6.17	&	P5	\\
ESO~215-09	&	597	&	122.0~~	&	93.0	&	4~~~	 &	1.967	&	H9	&	20.90	&	430	&	0.83	&	3.49	&	P6	\\
ESO~223-09	&	593	&	96.2	&	103.0~~	&	8.2	&	2.009	&	H10	&	19.21	&	1044~~	&	0.70	&	4.35	&	P7	\\
ESO~269-58	&	400	&	~~7.2	&	84.0	&	18~~~~	&	1.899	&	H11	&	19.10	&	636	&	0.63	&	4.72	&	P3	\\
ESO~320-14	&	654	&	~~2.5	&	61.3	&	18~~~~	&	1.738	&	H7	&	20.26	&	304	&	0.70	&	4.25	&	P7	\\
ESO~321-14	&	609	&	~~6.4	&	29.0	&	1.65	&	1.459	&	H8, H3	&	21.18	&	522	&	0.27	&	2.59	&	P6	\\
ESO~324-24	&	526	&	52.1	&	113.0~~	&	8.2	&	2.050	&	H10	&	20.43	&	630	&	0.93	&	3.13	&	P6	\\
ESO~325-11	&	550	&	25.4	&	77.0	&	8.2	&	1.880	&	H10	&	20.93	&	775	&	0.35	&	3.13	&	P3	\\
ESO~349-31	&	229	&	~~2.7	&	31.0	&	8.2	&	1.453	&	H10	&	21.51	&	323	&	0.58	&	3.21	&	P3	\\
ESO~379-07	&	644	&	~~5.2	&	40.0	&	~~1.65	&	1.600	&	H8, H3	&	22.10	&	443	&	0.85	&	2.81	&	P6	\\
ESO~381-18	&	625	&	~~3.3	&	61.6	&	18~~~~	&	1.741	&	H7	&	20.29	&	224	&	0.70	&	4.77	&	P6, P5	\\
ESO~381-20	&	596	&	31.9	&	103.0~~	&	8.2	&	2.009	&	H10	&	20.99	&	891	&	0.32	&	2.89	&	P8	\\
ESO~384-16	&	504	&	~~1.5	&	41.0	&	1.2	&	1.612	&	H12	&	19.45	&	214	&	0.92	&	3.00	&	P6	\\
ESO~444-78	&	573	&	~~4.0	&	52.1	&	1.4	&	1.716	&	H1	&	20.46	&	475	&	0.41	&	3.00	&	P1	\\
ESO~444-84	&	583	&	21.1	&	75.0	&	4~~~ 	&	1.873	&	H8, H13	&	20.60	&	318	&	0.89	&	2.46	&	P1	\\
ESO~461-36	&	427	&	~~7.5	&	84.0	&	10.2~~	&	1.916	&	H7, H14	&	20.51	&	492	&	0.50	&	4.58	&	P7	\\
GR~8	&	214	&	~~7.8	&	39.2	&	1.4	&	1.592	&	H1	&	21.36	&	152	&	0.80	&	3.70	&	P4	\\
Ho~II	&	156	&	267.0~~	&	72.0	&	5.2	&	1.854	&	H15	&	19.66	&	1348~~	&	1.00	&	2.25	&	P9	\\
IC~3104	&	429	&	10.3	&	63.0	&	18~~~~	&	1.753	&	H8	&	18.85	&	468	&	0.45	&	3.42	&	P3	\\
IC~4247	&	419	&	~~3.4	&	49.0	&	18~~~~	&	1.608	&	H16	&	18.77	&	370	&	0.33	&	5.30	&	P2, P5	\\
IC~4316	&	576	&	~~2.1	&	32.8	&	~~1.65	&	1.513	&	H16, H3	&	20.31	&	520	&	0.60	&	4.47	&	P5	\\
IC~4662	&	302	&	130.0~~	&	133.0~~	&	18~~~~	&	2.114	&	H8	&	17.40	&	242	&	0.73	&	5.69	&	P3	\\
IC~5152	&	122	&	97.2	&	100.0~~	&	18~~~~	&	1.983	&	H8	&	18.08	&	345	&	0.66	&	5.22	&	P3	\\
KK98~17	&	156	&	~~1.0	&	53.0	&	10.2~~	&	1.705	&	H14	&	21.67	&	470	&	0.31	&	2.98	&	P10	\\
KK98~182	&	613	&	~~2.1	&	24.0	&	7.9	&	1.316	&	H14	&	21.28	&	400	&	0.68	&	3.07	&	P3	\\
KK98~200	&	494	&	~~1.7	&	26.5	&	~~1.65	&	1.421	&	H1, H3	&	20.51	&	179	&	0.70	&	5.17	&	P7	\\
KK98~230	&	~~63	 &	~~2.6	&	25.9	&	~~1.65	&	1.411	&	H1, H17	&	22.57	&	140	&	0.95	&	1.76	&	P11	\\
KKH~086	&	287	&	~~0.5	&	20.6	&	1.4	&	1.310	&	H1 	&	21.56	&	181	&	0.61	&	1.98	&	P3	\\
KKH~098	&	$-$132~~	&	4.1	&	31.5	&	~~1.65	&	1.498	&	H1, H3	&	21.39	&	184	&	0.59	&	2.68	&	P10	\\
Mrk~178	&	250	&	~~3.0	&	44.8	&	1.4	&	1.650	&	H1	&	19.15	&	228	&	0.50	&	5.56	&	P4	\\
NGC~1311	&	568	&	14.6	&	105.0~~	&	18~~~~	&	2.005	&	H8	&	19.15	&	586	&	0.40	&	5.36	&	P12	\\
NGC~1569	&	$-$86	&	84.0	&	123.8~~	&	5.2	&	2.092	&	H18	&	16.83	&	271	&	0.55	&	6.13	&	P4	\\
NGC~2915	&	468	&	145.0~~	&	163.0~~	&	6.6	&	2.211	&	H19	&	17.49	&	306	&	0.50	&	7.67	&	P5	\\
NGC~3077	&	$-$20	&	256.0~~	&	157.3~~	&	5.2	&	2.196	&	H18	&	17.28	&	593	&	0.70	&	4.50	&	P9	\\
NGC~3738	&	225	&	22.0	&	122.0~~	&	~~8.24	&	2.084	&	H6	&	18.41	&	437	&	0.70	&	5.08	&	P4	\\
NGC~4163	&	164	&	~~9.6	&	38.0	&	4.1	&	1.574	&	H5, H20	&	19.29	&	223	&	0.70	&	5.18	&	P4	\\
NGC~4214	&	293	&	319.8~~	&	89.8	&	2.6	&	1.952	& H5, H18	&	17.63	&	491	&	0.50	&	4.46	&	P9	\\
NGC~5408	&	506	&	65.5	&	123.0~~	&	8.2	&	2.087	&	H10	&	18.88	&	456	&	0.50	&	7.03	&	P5	\\
NGC~6822	&	$-$55	&	2266.0~~~~	&	115.0~~	&	1.9	&	2.061	&	H21, H22	&	19.55	&	239	&	0.80	&	1.87	&	P9	\\
Peg~DIG	&	$-$183~~	&	28.1	&	38.6	&	5.3	&	1.577	&	H23	&	20.93	&	333	&	0.55	&	2.08	&	P10	\\
Sex~A	&	324	&	168.0~~	&	64.0	&	~~1.12	&	1.806	&	H24	&	21.09	&	362	&	0.95	&	3.03	&	P3	\\
Sex~B	&	301	&	72.9	&	56.0	&	1.4	&	1.747	&	H1	&	20.57	&	255	&	0.87	&	3.53	&	P6	\\
UGC~0685	&	156	&	13.4	&	83.0	&	~~1.65	&	1.919	&	H25, H3	&	19.72	&	336	&	0.70	&	5.07	&	P12	\\
UGC~3755	&	315	&	~~6.8	&	50.6~~	&	1.4	&	1.703	&	H1	&	19.81	&	808	&	0.50	&	3.72	&	P2	\\
UGC~4115	&	343	&	21.0	&	106.0~~	&	~~1.65	&	2.025	&	H5, H3	&	20.24	&	803	&	0.40	&	3.74	&	P4	\\
UGC~4483	&	156	&	13.6	&	50.6	&	1.4	&	1.703	&	H1	&	20.70	&	306	&	0.55	&	2.81	&	P1	\\
UGC~6456	&	$-$94	&	10.1	&	52.0	&	~~1.65	&	1.716	&	H3	&	20.16	&	248	&	0.70	&	5.08	&	P4	\\
UGC~7605	&	310	&	~~5.7	&	43.7	&	1.4	&	1.640	&	H1	&	20.75	&	402	&	0.67	&	3.91	&	P11	\\
UGC~8508	&	~~56	&	18.3	&	65.0	&	~~1.65	&	1.813	&	H3	&	19.95	&	234	&	0.55	&	4.64	&	P13	\\
UGC~8833	&	227	&	~~6.0	&	42.8	&	1.4	&	1.631	&	H1	&	20.94	&	274	&	0.77	&	4.05	&	P2	\\
UGCA~092	&	$-$95	&	104.7~~	&	73.0	&	~~1.65	&	1.863	&	H5, H3	&	20.53	&	635	&	0.50	&	2.40	&	P4	\\
UGCA~438	&	62	&	15.0	&	35.0	&	8.2	&	1.514	&	H26	&	20.48	&	283	&	0.90	&	3.97	&	P3	\\
WLM	&	$-$122~~~~	&	292.0~~	&	81.0	&	~~1.12	&	1.909	&	H24	&	21.28	&	437	&	0.40	&	3.06	&	P7	\\

\hline
\end{longtable}

\tablefoot{
(1) Name of galaxy;
(2) Heliocentric radial velocity defined by HI;
(3) Integrated HI flux;
(4) Apparent full width of 21 cm line at 20\% of peak;
(5) FWHM of instrumental profile;
(6) Logarithm of the 21 cm line width at 20\% of peak, corrected for instrumental broadening and redshift;
(7) Source of HI data;
(8) Central surface brightness of sech model, corrected for extinction and redshift;
(9) Scale length of sech model in parsecs;
(10) Adopted ratio of minor to major axes of isophotes;
(11) Ratio of radius of limiting isophote to scale length of sech;
(12) Source of surface photometry.
}

\tablebib{
(H1) \citet{huc03a};
(H2) \citet{beg03a};
(H3) \citet{beg08a};
(H4) \citet{spr05a};
(H5) \citet{huc86a};
(H6) \citet{sti02a};
(H7) \citet{mey04a};
(H8) \citet{kor04a};
(H9) \citet{war04a};
(H10) \citet{cot97a};
(H11) \citet{ban99a};
(H12) \citet{bea06a};
(H13) \citet{blo02a};
(H14) \citet{huc00a};
(H15) \citet{bur02a};
(H16) \citet{min03a};
(H17) \citet{beg06a};
(H18) \citet{wal08a};
(H19) \citet{meu96a};
(H20) \citet{swa02a};
(H21) \citet{blo06a};
(H22) \citet{wel03a};
(H23) \citet{kni09a};
(H24) \citet{bar04a};
(H25) \citet{gio93a}
(H26) \citet{lon78a};
(P1) CFHT 2004:  \citet{fin10a};  
(P2) SPM 2005:  \citet{fin10a};  
(P3) IRSF 2006:  \citet{fin10a};  
(P4) CFHT 2002:  \citet{vad05a};  
(P5) CTIO 2008:  this paper;  
(P6) IRSF 2005:  \citet{fin10a};  
(P7) CTIO 2007:  \citet{vad08a};  
(P8) CTIO 2006:  \citet{fin10a};  
(P9) 2MASS:  \citet{vad08a};  
(P10) CFHT 2005:  \citet{fin10a};  
(P11) CFHT 2006:  \citet{fin10a};  
(P12) ESO 2008:  this paper;  
(P13) SPM 2002:  \citet{vad05a}.  
}

}



\longtab{5}{
\begin{longtable}{lccccccccc}
\caption{\label{tbl_sample_derived}
Sample for Analysis:  Derived Global Properties
}\\
\hline
\hline
\noalign{\smallskip}

Galaxy &
Weight &
$M_{Ks}$	&
$L_{burst} / L_{sech}$ &
$\log M_{gas}$	&
$\log M_{stars}$ &
$\log M_{bary}$	 &
Gas Fraction &
$\log P$ &
$M / L \, \rm (virial)$
\\

& 
& 
(mag) & 
& 
($M_\odot$) & 
($M_\odot$) & 
($M_\odot$) &
&
($M_\odot \, \rm pc^{-1}$) &
($M_\odot / L_\odot$)
\\
 
(1) & 
(2) & 
(3) & 
(4) & 
(5) & 
(6) &
(7) &
(8) &
(9) &
(10)
\\

\hline

\endfirsthead
\caption{continued.}\\
\hline
\hline

\noalign{\smallskip}

Galaxy &
Weight &
$M_{Ks}$	&
$L_{burst} / L_{sech}$ &
$\log M_{gas}$	&
$\log M_{stars}$ &
$\log M_{bary}$	 &
Gas Fraction &
$\log P$ &
$M / L \, \rm (virial)$
\\

& 
& 
(mag) & 
& 
($M_\odot$) & 
($M_\odot$) & 
($M_\odot$) &
&
($M_\odot \, \rm pc^{-1}$) &
($M_\odot / L_\odot$)
\\
 
(1) & 
(2) & 
(3) & 
(4) & 
(5) & 
(6) &
(7) &
(8) &
(9) &
(10)
\\

\hline
\noalign{\smallskip}

\endhead
\hline
\endfoot

\noalign{\smallskip}

Cam~B	&	0	&	$-$14.51	&	0.010	&	7.213	&	7.077	&	7.451	&	0.578	&	4.842	&	2.95	\\
CGCG~087-33	&	1	&	$-$17.15	&	0.000	&	7.722	&	8.131	&	8.274	&	0.281	&	5.548	&	0.88	\\
DDO~006	&	1	&	$-$14.62	&	0.000	&	7.049	&	7.122	&	7.388	&	0.458	&	4.632	&	5.28	\\
DDO~047	&	0	&	$-$16.17	&	0.000	&	9.100	&	7.738	&	9.119	&	0.958	&	6.374	&	5.65	\\
DDO~099	&	1	&	$-$15.52	&	0.017	&	8.012	&	7.481	&	8.124	&	0.772	&	5.530	&	3.98	\\
DDO~167	&	1	&	$-$14.66	&	0.000	&	7.377	&	7.137	&	7.575	&	0.635	&	5.047	&	2.30	\\
DDO~168	&	0	&	$-$17.14	&	0.000	&	8.622	&	8.130	&	8.743	&	0.756	&	5.878	&	2.95	\\
DDO~181	&	1	&	$-$15.82	&	0.003	&	7.544	&	7.602	&	7.875	&	0.467	&	5.324	&	1.36	\\
DDO~187	&	1	&	$-$14.24	&	0.000	&	7.259	&	6.969	&	7.439	&	0.661	&	4.982	&	3.33	\\
DDO~190	&	1	&	$-$16.53	&	0.000	&	7.807	&	7.884	&	8.148	&	0.456	&	5.738	&	0.42	\\
DDO~226	&	1	&	$-$16.55	&	0.000	&	7.641	&	7.891	&	8.085	&	0.360	&	5.093	&	5.27	\\
ESO~059-01	&	1	&	$-$17.09	&	0.000	&	8.038	&	8.109	&	8.376	&	0.459	&	5.786	&	1.04	\\
ESO~121-20	&	1	&	$-$16.29	&	0.000	&	8.198	&	7.787	&	8.340	&	0.720	&	5.756	&	1.90	\\
ESO~137-18	&	1	&	$-$19.56	&	0.059	&	8.660	&	9.095	&	9.231	&	0.269	&	6.369	&	0.64	\\
ESO~215-09	&	0	&	$-$16.29	&	0.000	&	9.013	&	7.788	&	9.038	&	0.944	&	6.404	&	2.42	\\
ESO~223-09	&	0	&	$-$19.72	&	0.000	&	9.080	&	9.162	&	9.424	&	0.453	&	6.405	&	0.68	\\
ESO~269-58	&	1	&	$-$18.64	&	0.258	&	7.482	&	8.727	&	8.751	&	0.054	&	5.948	&	0.50	\\
ESO~320-14	&	1	&	$-$15.99	&	0.000	&	7.447	&	7.667	&	7.872	&	0.376	&	5.389	&	0.81	\\
ESO~321-14	&	1	&	$-$15.21	&	0.012	&	7.311	&	7.357	&	7.636	&	0.474	&	4.918	&	2.18	\\
ESO~324-24	&	1	&	$-$17.71	&	0.000	&	8.337	&	8.356	&	8.648	&	0.489	&	5.849	&	1.78	\\
ESO~325-11	&	1	&	$-$16.60	&	0.000	&	7.954	&	7.912	&	8.234	&	0.524	&	5.345	&	4.57	\\
ESO~349-31	&	1	&	$-$14.68	&	0.000	&	6.917	&	7.142	&	7.345	&	0.373	&	4.835	&	1.35	\\
ESO~379-07	&	1	&	$-$15.18	&	0.014	&	7.614	&	7.345	&	7.801	&	0.650	&	5.155	&	2.44	\\
ESO~381-18	&	1	&	$-$15.29	&	0.000	&	7.444	&	7.390	&	7.719	&	0.531	&	5.369	&	0.84	\\
ESO~381-20	&	1	&	$-$16.75	&	0.000	&	8.455	&	7.971	&	8.579	&	0.753	&	5.629	&	7.68	\\
ESO~384-16	&	1	&	$-$16.34	&	0.045	&	6.962	&	7.807	&	7.865	&	0.125	&	5.535	&	0.20	\\
ESO~444-78	&	1	&	$-$16.18	&	0.000	&	7.518	&	7.745	&	7.948	&	0.372	&	5.271	&	1.56	\\
ESO~444-84	&	0	&	$-$16.01	&	0.000	&	8.140	&	7.676	&	8.269	&	0.745	&	5.767	&	1.30	\\
ESO~461-36	&	1	&	$-$16.42	&	0.000	&	8.160	&	7.842	&	8.330	&	0.675	&	5.638	&	2.41	\\
GR~8	&	1	&	$-$13.53	&	0.000	&	6.972	&	6.684	&	7.153	&	0.660	&	4.970	&	1.28	\\
Ho~II	&	0	&	$-$20.21	&	0.000	&	8.969	&	9.358	&	9.507	&	0.290	&	6.377	&	0.46	\\
IC~3104	&	1	&	$-$17.86	&	0.000	&	7.219	&	8.416	&	8.442	&	0.060	&	5.772	&	0.36	\\
IC~4247	&	1	&	$-$17.06	&	0.000	&	7.403	&	8.097	&	8.177	&	0.168	&	5.608	&	0.30	\\
IC~4316	&	1	&	$-$16.94	&	0.109	&	7.031	&	8.050	&	8.089	&	0.087	&	5.374	&	0.51	\\
IC~4662	&	1	&	$-$18.41	&	0.000	&	8.347	&	8.634	&	8.815	&	0.341	&	6.431	&	0.17	\\
IC~5152	&	1	&	$-$18.38	&	0.077	&	8.039	&	8.625	&	8.725	&	0.206	&	6.187	&	0.24	\\
KK98~17	&	0	&	$-$14.64	&	0.005	&	6.868	&	7.129	&	7.319	&	0.354	&	4.647	&	6.11	\\
KK98~182	&	0	&	$-$15.54	&	0.000	&	7.342	&	7.487	&	7.722	&	0.417	&	5.120	&	0.62	\\
KK98~200	&	1	&	$-$14.59	&	0.000	&	6.992	&	7.109	&	7.356	&	0.433	&	5.103	&	0.41	\\
KK98~230	&	0	&	$-$12.34	&	0.015	&	6.508	&	6.207	&	6.684	&	0.667	&	4.537	&	1.93	\\
KKH~086	&	0	&	$-$13.42	&	0.000	&	6.039	&	6.638	&	6.736	&	0.201	&	4.478	&	0.89	\\
KKH~098	&	1	&	$-$13.59	&	0.000	&	6.870	&	6.706	&	7.097	&	0.593	&	4.832	&	1.36	\\
Mrk~178	&	1	&	$-$16.12	&	0.000	&	7.128	&	7.720	&	7.819	&	0.204	&	5.461	&	0.32	\\
NGC~1311	&	1	&	$-$17.92	&	0.194	&	8.135	&	8.442	&	8.616	&	0.330	&	5.848	&	1.11	\\
NGC~1569	&	1	&	$-$18.91	&	0.348	&	8.272	&	8.837	&	8.941	&	0.214	&	6.509	&	0.12	\\
NGC~2915	&	0	&	$-$18.41	&	0.000	&	8.777	&	8.636	&	9.013	&	0.580	&	6.527	&	0.35	\\
NGC~3077	&	0	&	$-$20.43	&	0.107	&	9.044	&	9.443	&	9.589	&	0.285	&	6.816	&	0.20	\\
NGC~3738	&	1	&	$-$18.63	&	0.204	&	8.169	&	8.725	&	8.831	&	0.218	&	6.191	&	0.40	\\
NGC~4163	&	1	&	$-$16.28	&	0.000	&	7.383	&	7.785	&	7.930	&	0.284	&	5.582	&	0.21	\\
NGC~4214	&	0	&	$-$19.29	&	0.016	&	8.969	&	8.989	&	9.280	&	0.488	&	6.589	&	0.19	\\
NGC~5408	&	1	&	$-$17.89	&	0.000	&	8.703	&	8.428	&	8.888	&	0.653	&	6.229	&	0.88	\\
NGC~6822	&	0	&	$-$16.32	&	0.000	&	8.191	&	7.802	&	8.339	&	0.710	&	5.960	&	0.95	\\
Peg~DIG	&	0	&	$-$15.25	&	0.000	&	6.835	&	7.374	&	7.484	&	0.224	&	4.962	&	1.20	\\
Sex~A	&	1	&	$-$15.87	&	0.000	&	8.040	&	7.622	&	8.180	&	0.723	&	5.622	&	1.56	\\
Sex~B	&	1	&	$-$15.53	&	0.000	&	7.673	&	7.485	&	7.890	&	0.606	&	5.484	&	0.90	\\
UGC~0685	&	1	&	$-$16.75	&	0.000	&	7.975	&	7.973	&	8.275	&	0.501	&	5.748	&	0.84	\\
UGC~3755	&	0	&	$-$18.20	&	0.000	&	8.082	&	8.552	&	8.679	&	0.253	&	5.771	&	0.68	\\
UGC~4115	&	1	&	$-$17.51	&	0.000	&	8.606	&	8.277	&	8.773	&	0.681	&	5.868	&	3.23	\\
UGC~4483	&	1	&	$-$15.31	&	0.000	&	7.734	&	7.395	&	7.898	&	0.686	&	5.412	&	1.40	\\
UGC~6456	&	1	&	$-$15.66	&	0.000	&	7.785	&	7.535	&	7.979	&	0.640	&	5.584	&	0.69	\\
UGC~7605	&	1	&	$-$16.06	&	0.000	&	7.532	&	7.697	&	7.923	&	0.406	&	5.319	&	1.00	\\
UGC~8508	&	1	&	$-$15.46	&	0.000	&	7.559	&	7.457	&	7.812	&	0.558	&	5.443	&	0.97	\\
UGC~8833	&	1	&	$-$15.19	&	0.000	&	7.244	&	7.349	&	7.601	&	0.440	&	5.163	&	1.01	\\
UGCA~092	&	0	&	$-$16.95	&	0.000	&	8.520	&	8.054	&	8.648	&	0.745	&	5.845	&	2.11	\\
UGCA~438	&	1	&	$-$15.89	&	0.000	&	7.350	&	7.629	&	7.812	&	0.345	&	5.361	&	0.40	\\
WLM 	&	1	&	$-$15.15	&	0.000	&	7.898	&	7.331	&	8.002	&	0.787	&	5.361	&	6.02	\\

\hline
\end{longtable}

\tablefoot{
(1) Name of galaxy;
(2) Weight during fitting;
(3) Absolute magnitude of sech model in $K_s$;
(4) Luminosity of burst relative to luminosity of sech model;
(5) Logarithm of the gas mass;
(6) Logarithm of the mass of stars in the sech model, computed assuming a fixed mass-to-light ratio of 0.883 in $K_s$;
(7) Logarithm of the baryonic mass (sum of gaseous and stellar masses);
(8) Gas fraction (for a fixed mass-to-light ratio of 0.883 in $K_s$);
(9) Logarithm of the potential defined by the ratio of the baryonic mass to the scale length of the sech model;
(10) Mass-to-light ratio of the stars in $K_s$ as indicated by the deviation of the surface brightness from the norm for a virialized system.
}

}



\end{document}